\documentclass[aps,prl,twocolumn,superscriptaddress,footinbib,longbibliography]{revtex4-1} 

\usepackage{longtable}
\usepackage{morefloats}
\usepackage[dvips]{graphicx}
\usepackage{color}
\usepackage{epsfig,graphicx,amsfonts,amsbsy}
\usepackage{amsmath,amsfonts,amsthm,amssymb}
\usepackage{appendix}
\usepackage{makeidx}
\usepackage{url}
\usepackage{verbatim}
\usepackage[bookmarksnumbered,pdfpagelabels=true,plainpages=false,colorlinks=true,linkcolor=blue,citecolor=blue,urlcolor=blue]{hyperref}
\usepackage[rightcaption]{sidecap}
\usepackage{array}
\usepackage{multirow}
\usepackage{tabularx}
\usepackage{braket} 
\usepackage{dsfont}
\usepackage{bbold}
\usepackage{etoolbox}
\usepackage{booktabs}

\AtBeginDocument{%
    \newwrite\bibnotes
    \def\bibnotesext{Notes.bib}
    \immediate\openout\bibnotes=\jobname\bibnotesext
    \immediate\write\bibnotes{@CONTROL{REVTEX41Control}}
    \immediate\write\bibnotes{@CONTROL{%
    apsrev41Control,author="08",editor="1",pages="0",title="0",year="1"}}
     \if@filesw
     \immediate\write\@auxout{\string\citation{apsrev41Control}}%
    \fi
}%

\newcommand{\bs}[1]{\boldsymbol{#1}}
\newcommand{\ssf}[1]{\mathsf{#1}}

\def\half{{\tfrac{1}{2}}}

\def\BD{{\bs{D}}}

\def\BD{{\bs{D}}}

\newcommand{\pt}{\partial}
\newcommand{\mb}{\mathbf}
\newcommand{\mc}{\mathcal}
\newcommand{\Jint}{J_{\scalebox{0.6}{int}}}

\def\muB{{ \mu_{\mbox{\tiny B}} }}

\def\CalO{{\mathcal{O}}}

\newcommand{\EC}{\mathcal{E}_{C_{2v}}}
\newcommand{\ED}{\mathcal{E}_{D_{2d}}}
\newcommand{\EB}{\mathcal{E}_{\scalebox{0.7}{cub}}}
\newcommand{\EIF}{\mathcal{E}_{\scalebox{0.7}{int}}}

\newcommand{\Edm}{\mathcal{E}_{\scalebox{0.7}{DM}}}
\newcommand{\Hint}{H_{\scalebox{0.7}{int}}}
\newcommand{\Hosc}{H_{\scalebox{0.7}{osc}}} 
\newcommand{\wres}{\omega_{\scalebox{0.7}{res}}} 
\newcommand{\Mfm}{n^z_{\scalebox{0.7}{FM}}} 
 
\newcommand{\Eint}{E_{\scalebox{0.7}{int}}}
\newcommand{\kres}{k_{\scalebox{0.7}{res}}}

\newcommand{\egap}{\varepsilon_{\scalebox{0.7}{gap}}}

\def\SS{{\ssf{S}}}

\def\Hsw{{H_{\mbox{\tiny SW}}}}

\begin{document}

\title{Spin Wave Radiation by a Topological Charge Dipole}

\author{Sebasti{\'a}n A. D{\'i}az}
\affiliation{Department of Physics, University of Basel, Klingelbergstrasse 82, CH-4056 Basel, Switzerland}
\author{Tomoki Hirosawa}
\affiliation{Department of Physics, University of Tokyo, Bunkyo, Tokyo 113-0033, Japan}
\author{Daniel Loss}
\affiliation{Department of Physics, University of Basel, Klingelbergstrasse 82, CH-4056 Basel, Switzerland}
\author{Christina Psaroudaki}
\affiliation{Department of Physics, California Institute of Technology, Pasadena, CA 91125, USA}
\affiliation{Institute for Theoretical Physics, University of Cologne, D-50937 Cologne, Germany}

\date{\today}
	
\begin{abstract}
The use of spin waves (SWs) as data carriers in spintronic and magnonic logic devices offers operation at low power consumption, free of Joule heating. Nevertheless, the controlled emission and propagation of SWs in magnetic materials remains a significant challenge. Here, we propose that skyrmion-antiskyrmion bilayers form topological charge dipoles and act as efficient sub-100 nm SW emitters when excited by in-plane ac magnetic fields. The propagating SWs have a preferred radiation direction, with clear dipole signatures in their radiation pattern, suggesting that the bilayer forms a SW antenna. Bilayers with the same topological charge radiate SWs with spiral and antispiral spatial profiles, enlarging the class of SW patterns. We demonstrate that the characteristics of the emitted SWs are linked to the topology of the source, allowing for full control of the SW features, including their amplitude, preferred direction of propagation, and wavelength. 
\end{abstract}

\maketitle


\section{Introduction}

Magnetic skyrmions, particle-like textures in quasi-two-dimensional (2D) systems, are promising elements in future magnetic memory devices \cite{fernandezpacheco2017}, with complex dynamics governed by topology \cite{doi:10.1063/1.5048972}. In systems that lack inversion symmetry, isotropic Dzyaloshinskii-Moriya (DM) interactions energetically stabilize skyrmions \cite{BOGDANOV1994255}, while anisotropic DM interactions can stabilize antiskyrmions \cite{Hoffmann2017,Nayak2017} with opposite topological charge \cite{Koshibae2016,PhysRevB.97.134404,PhysRevB.93.064428}. This new class of skyrmions, along with more exotic fractional objects \cite{PhysRevB.91.224407}, broaden the family of topological magnetic particles and motivate the fabrication of new materials where rich topological phenomena are expected. 

The topology of these structures gives rise to fascinating properties, including the topological Hall effect in charge transport \cite{PhysRevLett.102.186602,PhysRevLett.102.186601,PhysRevLett.110.117202,Schulz2012} and the skyrmion Hall effect \cite{Litzius2017,Jiang2017}, in which skyrmions are deflected in a direction transverse to the applied force. Skyrmions and antiskyrmions, with opposite topological charges, have opposite lateral deviations \cite{Leonov2017,Everschor_Sitte_2017}. When realized in bilayers, they lead to the absence of the skyrmion Hall effect \cite{PhysRevB.96.144412,Zhang2016,PhysRevB.94.064406} and result in  more reliable information carriers. 

The non-trivial skyrmion topology affects the properties of the surrounding magnons \cite{PhysRevB.90.094423,PhysRevLett.122.187203,1910.05214}, collective spin excitations, relevant for future magnetic logic and memory devices \cite{Khitun_2010,6922543} with tailored properties \cite{LENK2011107,Serga_2010,Yu2014,Yu2016}. SW currents can be used to transport and process information, free of Joule heating, a significant drawback of modern electronics. Controlled SW emission with nanoscale wavelength, required for the fabrication of miniature devices, has been the subject of intense scientific investigations \cite{Wintz2016,Maci2014,doi:10.1063/1.3631756,SLONCZEWSKI1996L1,Tsoi2000}. Notably, noncollinear spin structures, including skyrmions, allow for a controlled SW transmission \cite{Duerr_2011}, while skyrmion-hosting magnetic insulators are ideal materials for microwave technologies \cite{Garst_2017}. 

\begin{figure}[b!]
\centering
\includegraphics[width=1\columnwidth]{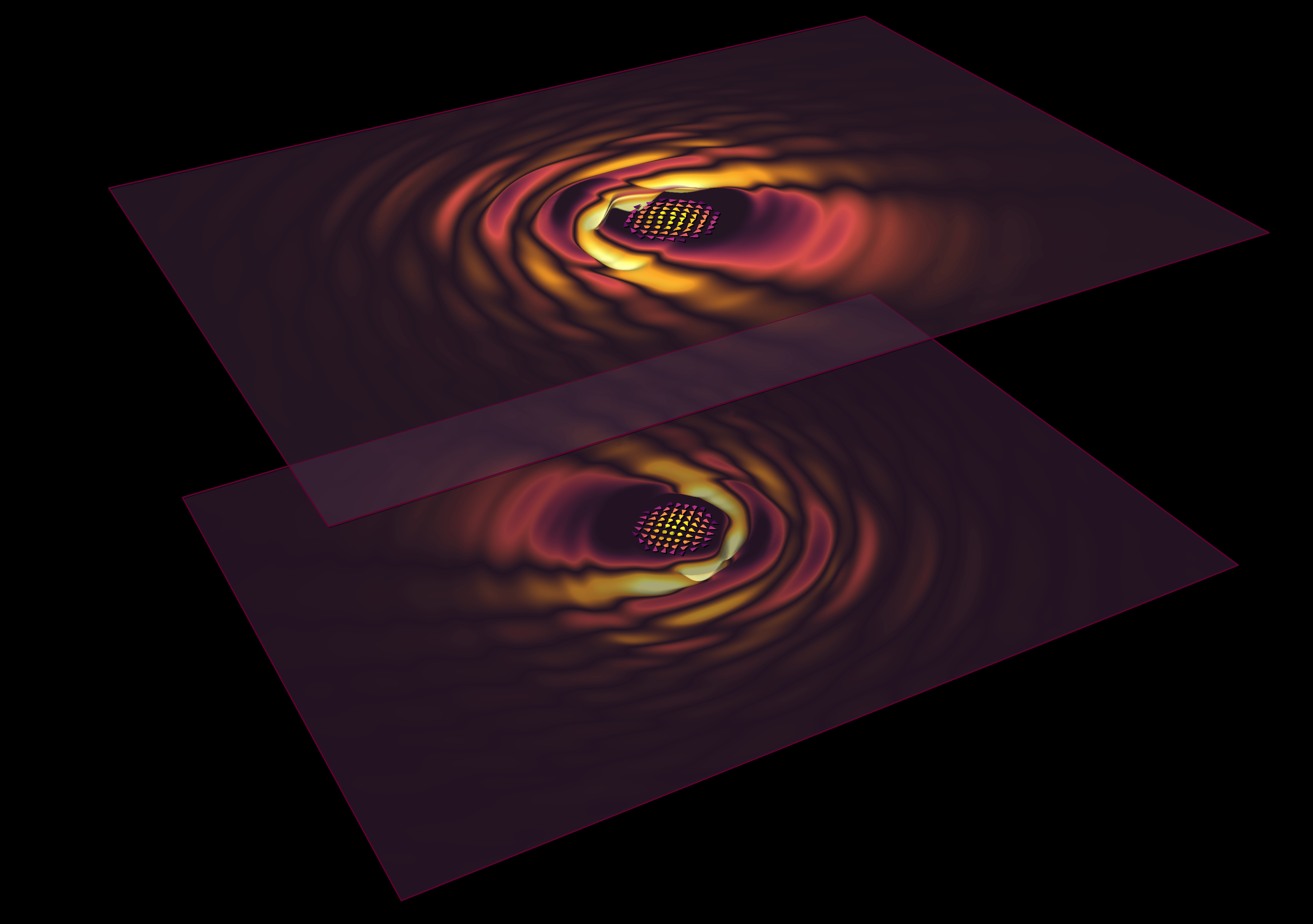}
\caption{Skyrmion-antiskyrmion bilayer in the presence of an in-plane microwave magnetic field (not shown). The emitted spin wave pattern has dipole signatures, suggesting that the bilayer forms a topological charge dipole that acts as an efficient spin-wave antenna. }
\label{fig:bilayer}
\end{figure}
\begin{figure*}[t!]
\centering
\includegraphics[scale=0.55]{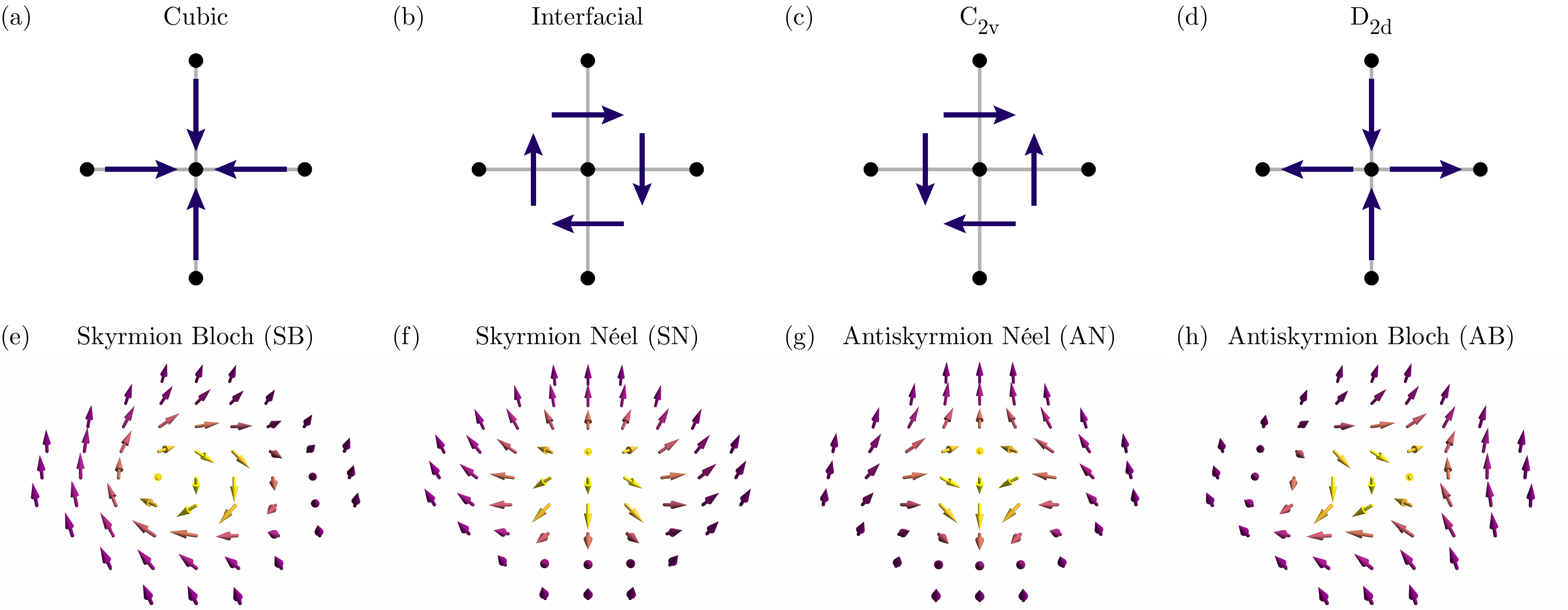}
\caption{Relation between crystal symmetry and spin configuration. (a)-(d) Sketch of the DM vectors (blue arrows) between n.n. sites of the lattice for the cubic, interfacial, $C_{2v}$ and $D_{2d}$ symmetry. (e)-(h) Spin field configuration, corresponding to different DM interactions. In all cases, the $z$-component of magnetization is aligned parallel to the external out-of-plane magnetic field. (e) A Bloch skyrmion (SB) with $\gamma=-\pi/2$ and $Q=-1$, (f) a N\'{e}el skyrmion (SN) with $\gamma=0$ and $Q=-1$, (g) A N\'{e}el antiskyrmion (AN) with $\gamma=\pi$ and $Q=1$, and (h) a Bloch antiskyrmion (AB) with $\gamma=\pi/2$ and $Q=1$.  }
\label{fig:DMI_types}
\end{figure*}

Here, by means of micromagnetic simulations, we consider a skyrmion-antiskyrmion bilayer in the presence of an in-plane ac magnetic field, which activates a counterclockwise (clockwise) rotation of the skyrmion (antiskyrmion) core. Spin waves emitted by the interacting gyrating cores have a preferred propagating direction, with dipole signatures in their radiation pattern, suggesting that the bilayer forms a topological charge dipole, which acts as a spin-wave antenna (see Fig.~\ref{fig:bilayer}). The far-field SW amplitude can be controlled by the interlayer coupling, and the corresponding wavelength is in the sub-100 nm regime. Spiral and antispiral SW patterns are obtained in bilayers with the same topological charge and different helicity, with a rotation that depends on the topology of the source. Bilayers with the same topological charge and helicity form a monopole source and emit radially symmetric waves. Our discoveries suggest that skyrmion-antiskyrmion bilayers excited by ac magnetic fields, form topological charge dipoles, which besides being of fundamental interest, can be used as efficient SWs emitters with enriched and controlled characteristics. 

\section{Results}

\textbf{Topological textures}. We consider the following spin-lattice Hamiltonian, defined on a 2D square lattice structure,
\begin{equation}\label{eq:SpinH}
H = - \half \sum_{\mb{r},i} (
J \mb{S}_\mb{r} \cdot \mb{S}_{\mb{r} \pm a \mb{e}_i} + D \mb{d}_{\pm \mb{e}_i} \cdot \mb{S}_\mb{r} \times \mb{S}_{\mb{r} \pm a \mb{e}_i} ) +H_z \,,
\end{equation}
where $\mb{S}_\mb{r}$ is a spin-$S$ operator at site $\mb{r}$, and $H_z=- \sum_\mb{r}  g\mu_B B S^z_\mb{r}$ is the Zeeman energy along the $z$ axis. $J$ and $D$ represent the exchange and DM \cite{DZYALOSHINSKY1958241,PhysRev.120.91} couplings respectively, $g$ is the g-factor, $\mu_B$ the Bohr magneton, $a$ is the lattice constant, and $\mb{e}_{x,y}$ are the unit vectors in the $x$ and $y$ directions respectively. The DM interaction, which satisfies $\mb{d}_{\mb{e}_i}=-\mb{d}_{-\mb{e}_i}$, is the result of spin-orbit coupling and lack of inversion symmetry. The specific form and orientation of the unit vector $\mb{d}_{\pm \mb{e}_i}$ of a given material, is for the most part dictated by its crystal symmetry, and determines the chiral spin configuration. 
\begin{table*}[]
\centering
\caption{Composite pairs of skyrmions and antiskyrmions. Type I includes all possible combinations of particles with opposite topological charge, Type II of the same charge and helicity, and Type III of the same charge and different helicity. We introduce $\tilde{Q}=Q_1+Q_2$, and $\tilde{\gamma}=\gamma_1-\gamma_2$.}
\label{tab:Types}
\begin{ruledtabular}
\begin{tabular}{c|c|c|c|c||c|c|c|c||c|c}
 & \multicolumn{4}{c}{Type I} & \multicolumn{4}{c}{Type II} & \multicolumn{2}{c}{Type III} \\ 
 \hline
Subtype & a & b & c & d & a & b & c & d & a & b \\
Composite Pair & SB-AN & SN-AB & SN-AN & SB-AB & SB-SB & SN-SN & AN-AN & AB-AB & SB-SN & AN-AB \\ 
($\tilde{Q}$,$\tilde{\gamma}$) & (0,$-3\pi/2$) & (0,$3\pi/2$) & (0,$-\pi$) & (0,$-\pi$) & ($-2$,0) & ($-2$,0) & (2,0) & (2,0) & ($-2$,$-\pi/2$) & (2,$3\pi/2$) \\ 
\end{tabular}
\end{ruledtabular}
\end{table*}

To get an insight on the type of magnetization textures supported by the model of Eq.~\eqref{eq:SpinH}, we consider the continuum limit in which $\mb{n}_\mb{r}=\mb{S}_\mb{r}/S$, with $S$ the total spin, turns into a field $\mb{n}(\mb{r})$, usually expressed in spherical parametrization $\mb{n}=[\sin \Theta \cos \Phi,  \sin \Theta \sin \Phi, \cos\Theta]$ (see Supplementary Note \hyperref[sec:Continuum]{3} for details). Topological textures are described by $\Phi(\mb{r})=\mu \phi+ \gamma$ and $\Theta(\mb{r}) = \Theta(\rho)$, with $\mb{r}=(\rho,\phi)$ the polar coordinate system, and $\gamma$ the helicity. These topological solutions are characterized by an integer topological charge $Q$,
\begin{align}
Q=\frac{1}{4 \pi} \int d\mb{r}~ \mb{n} \cdot (\pt_x \mb{n} \times \pt_y \mb{n}) \,,
\end{align}
which denotes the mapping from the 2D magnetic system in real space into the 3D spin space \cite{PhysRevLett.51.2250}. Under a choice a ferromagnetic background $\mb{n}_0 = (0,0,1)$, the topological charge $Q$ is related to the winding number $\mu$ as $Q=-\mu$ \cite{Hoffmann2017}. Depending on the crystal symmetry, both the value and the sign of $\mu$, $Q$, and $\gamma$ are uniquely defined. Here we consider four cases of crystal symmetry, namely cubic, interfacial, $C_{2v}$, and $D_{2d}$, depicted in Fig.~\ref{fig:DMI_types}, while in all cases we keep $J>0$, and $D>0$. The cubic symmetry stabilizes a Bloch skyrmion (SB), with $Q=-1$ and $\gamma=-\pi/2$, the interfacial a N\'{e}el skyrmion (SN) with $Q=-1$ and $\gamma=0$, the $D_{2d}$ a Bloch antiskyrmion (AB) with $Q=1$, and $\gamma=\pi/2$, and the $C_{2v}$ a N\'{e}el antiskyrmion (AN) with $Q=1$ and $\gamma=\pi$. In Fig.\ref{fig:DMI_types}-(a)-(d) we summarize the various types of DM vectors for each type of symmetry, and in \ref{fig:DMI_types}-(e)-(h) the resulting spin configurations. To make our considerations relevant to experimental studies we note that, noncentrosymmetric magnetic materials with cubic symmetry \cite{PhysRevB.82.052403,Yu2011} as well as magnetic thin films on nonmagnetic metals with strong spin orbit coupling, thus inducing interfacial DM interactions \cite{Heinze2011}, can host skyrmions. A double layer Fe on W(110) exhibits a $C_{2v}$ symmetry and has been suggested to support antiskyrmions \cite{Hoffmann2017}, which have been recently reported in acentric tetragonal Heusler compounds with $D_{2d}$ crystal symmetry \cite{Nayak2017}. 

\textbf{Magnetic Excitations}. In the following we discuss the energy spectrum of magnons supported by any of the textures depicted in Fig.~\ref{fig:DMI_types}-(e)-(f), obtained by a numerical diagonalization of the spin wave Hamiltonian $\Hsw$. In Supplementary Note \hyperref[sec:Magnons]{2} we provide an explicit, detailed construction of $\Hsw$ and information on the diagonalization procedure. Usual propagating SWs carry energy $\mc{E} =\egap + (\hbar^2/2 m) k^2$, where $\egap = g\mu_B B$ is the gap due to the magnetic field, $k$ is the radial momentum, and $m = \hbar^2/2 JS^2 a^2$ is the magnon mass. In addition, there exist a number of localized states, corresponding to deformations of the skyrmion \cite{PhysRevB.90.094423,PhysRevB.89.024415}. Of particular importance for the present study is the counterclockwise (CCW) mode for the skyrmion core, which has been experimentally measured in the skyrmion-hosting ferrimagnetic insulator Cu$_2$OSeO$_3$ in the GHz regime \cite{PhysRevLett.109.037603,Okamura2013}. The high-energy part of the magnon spectrum corresponds to bands that reside at THz frequencies \cite{PhysRevLett.113.157205}. 

From the obtained values of magnon energies around any of the textures of Fig.~\ref{fig:DMI_types}, we conclude that the magnon spectrum is insensitive to the choice of $Q$ and $\gamma$. We observe however, that the sense of gyration of localized deformations depends on the sign of $Q$. Local modes of the skyrmion with a CCW sense of gyration, correspond to clockwise (CW) modes for the antiskyrmion. This observation is confirmed by an analytical derivation of the magnon eigenvalue problem derived in the continuum model, given explicitly in Supplementary Note \hyperref[sec:Continuum]{3}. We numerically confirm the existence of a CCW mode for the skyrmion, with energy $\mc{E}_0 \approx \egap$, and of a CW mode for the antiskyrmion, at the same energy. These modes describe a rotation of the out-of-plane spin components around the (anti)skyrmion core in a (CW) CCW manner, and can be excited by an in-plane ac magnetic field \cite{Ogawa2015}. The remaining of the localized modes are depicted in Fig.~\ref{fig:MagnonSpec}, where we plot the energies of the 10 lowest-lying magnon modes as a function of the external magnetic field.


~\\
\textbf{Skyrmions in Bilayers}. We now proceed by considering a bilayer of magnetic materials, such as the one illustrated in Fig.~\ref{fig:bilayer}, where each layer hosts a texture with finite topological charge that can be any of the ones depicted in Fig.~\ref{fig:DMI_types}(e)-(h). Each magnetic layer is described by the model of Eq.~\eqref{eq:SpinH}, with corresponding DM vectors shown in Fig.~\ref{fig:DMI_types}(a)-(d). The various composite pairs are summarized in Table~\ref{tab:Types}, and are categorized in three different types. Type I includes all possible combinations of particles with opposite topological charge, Type II of the same charge and helicity, and Type III of the same charge and different helicity. As we demonstrate below, the dynamics of the topological particles, as well as the characteristics of the emitted SWs, strongly depend on the composite pair type. 

The two layers are coupled through a ferromagnetic interaction, $\Hint= -\Jint \sum_{\mb{r}} \mb{S}_{\mb{r}}^1 \cdot \mb{S}_{\mb{r}}^2$, with $\Jint>0$, the interlayer ferromagnetic coupling. $\Jint$ can be tuned experimentally by introducing a spacer between the two layers, such as a nonmagnetic insulating material \cite{Koshibae2016,Chen2013}. 
For reasons of simplicity, in all considered cases, the $J$ and $D$ couplings in both layers have the same strength, thus the skyrmion and antiskyrmion have the same size. The two particles interact via a potential of the form $\Eint (R_0)= \Jint \int d\mb{r} [1-\mb{n}_1(\mb{r}-\mb{R}_1) \cdot \mb{n}_2(\mb{r}-\mb{R}_2)] d\mb{r}$, where $\mb{R}_i$ are the collective coordinates of position for each particle \cite{PhysRevX.7.041045,PhysRevLett.120.237203}, and $R_0=\vert \mb{R}_1 -\mb{R}_2\vert$ \cite{Koshibae2017}. In Supplementary Note \hyperref[sec:Continuum]{3}, we calculate $\Eint$ for all bilayer types based on the continuum model. From Fig.~\ref{fig:inter}, where we present $\Eint$ as a function of $R_0$, we conclude that $\Eint$ is an even function in $R_0$ and depends on both $Q$ and $\gamma$. 

\begin{figure*}[t]
\centering
\includegraphics[scale=0.55]{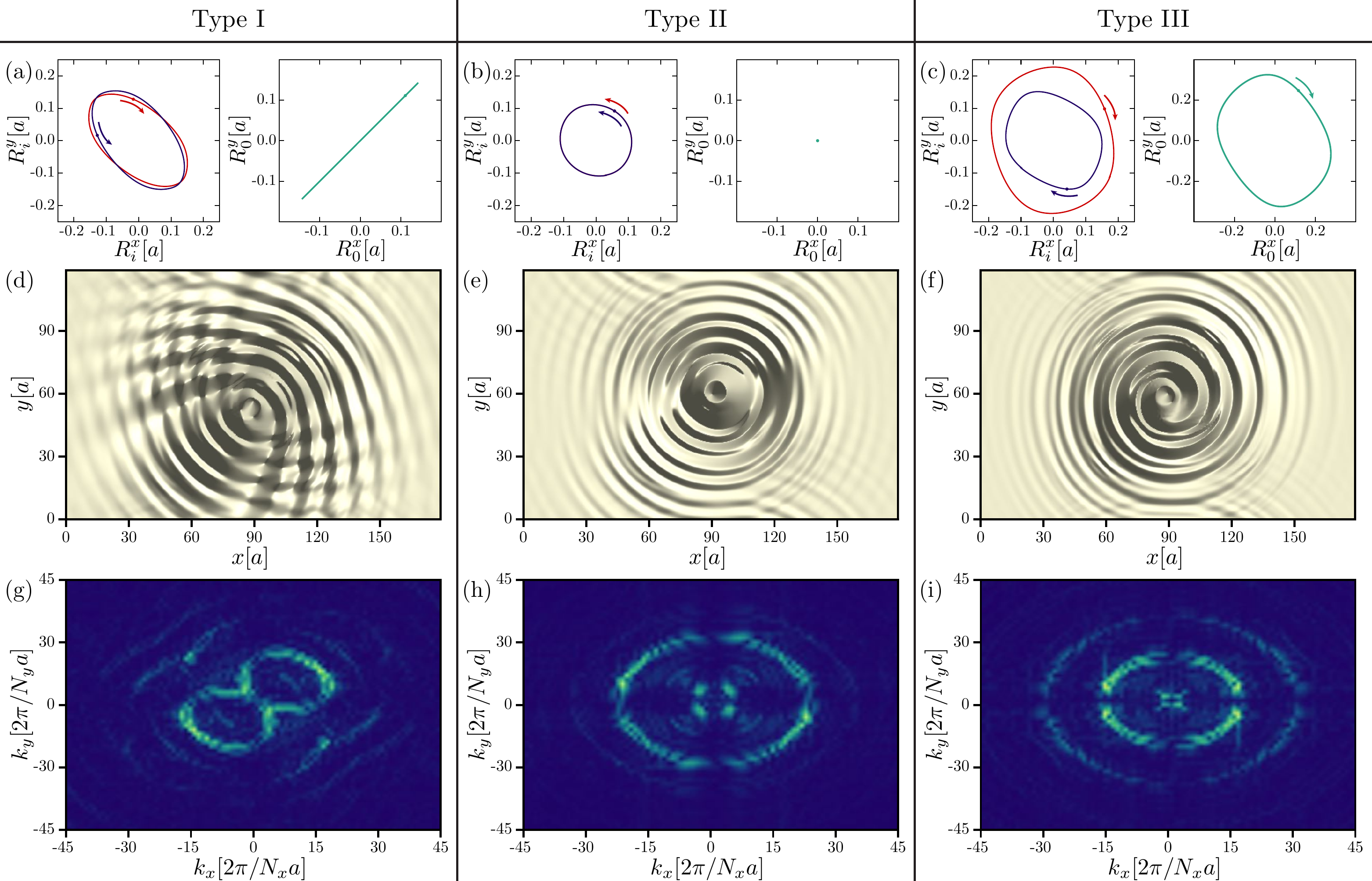}
\caption{Relation between the dynamics of the collective coordinate of position $\mb{R}_i(t)$ and the features of the radiated SWs. In all cases we use $J/D=1$, $b=0.6$ ($B=0.6 JS/g\mu_B$), and $\omega=0.6$ ($\tilde{\omega}=0.6 JS/\hbar$). (a)-(c) Blue lines denote the path of $\mb{R}_i$ for the texture in layer 1, and red lines for layer 2, while arrows indicate the sense of gyration. The green lines denote the path of the charge separation distance $\mb{R}_0 = \mb{R}_1 - \mb{R}_2$. (d)-(f) Snapshots of the perpendicular magnetization deviation $\delta n^z_{\mb{r}}(t)$ and (g)-(i) the radiation pattern $\delta n^z_{\mb{k}}$, defined here as the absolute value of the Fourier Transform of  $\delta n^z_{\mb{r}}(t)$, averaged over one period of time.}
\label{fig:SW_Types}
\end{figure*}
 

~\\
\textbf{Topological Charge Dipole as a SW antenna}. We now turn to the main task of this paper and calculate the magnetization dynamics of the bilayer by numerically solving the Landau-Lifshitz-Gilbert (LLG) equation \cite{TATARA2008213} to evaluate the time evolution of $\mb{n}^i_{\mb{r}}(t)=\mb{S}^i_{\mb{r}}(t)/M_s$, for each of the $i=1,2$ layers, with $M_s$ the saturation magnetization. The dynamics is governed by the total Hamiltonian, $H= H_1 + H_2 + \Hint + \Hosc$, with $\Hosc= -\sum_{\mb{r}} g\mu_B B_0 \cos(\omega t) (n_{\mb{r}}^{x,1}+ n_{\mb{r}}^{x,2})$ describing the presence of a time-periodic in-plane magnetic field of frequency $\omega$. We consider two coupled monolayers of a finite lattice of $180 \times 120$ sites in the $xy$ plane, and periodic boundary conditions. The simulations were performed using $J=1$, Gilbert damping $\alpha=0.08$, $b_0 = g\mu_B B_0/JS=0.1$, and unless explicitly stated, $D=1$.  Time $t$, frequency $\omega$, and space $\mb{r}$ are given in dimensionless units. Physical units are restored as $\tilde{t}= \hbar t /JS$, $\tilde{\mb{r}}=\mb{r} a$, and $\tilde{\omega}=JS \omega/\hbar$. The full LLG simulation is up to $10^{4}$ time steps. 
We numerically verify that $\Hosc$ activates the CCW (CW) mode for a skyrmion (antiskyrmion), signaled by a resonance peak at $\omega \simeq b$ (see \hyperref[Meth:MicroMagn]{Methods} for details on the micromagnetic simulation). For $\Jint=0$, $\omega=0.6$, and $b=0.6$, each of the uncoupled layers emits radially symmetric SWs with a structure similar to the one depicted in Eq.~\ref{fig:SW_Types}-(e). 
\begin{figure*}[t]
\centering
\includegraphics[scale=0.55]{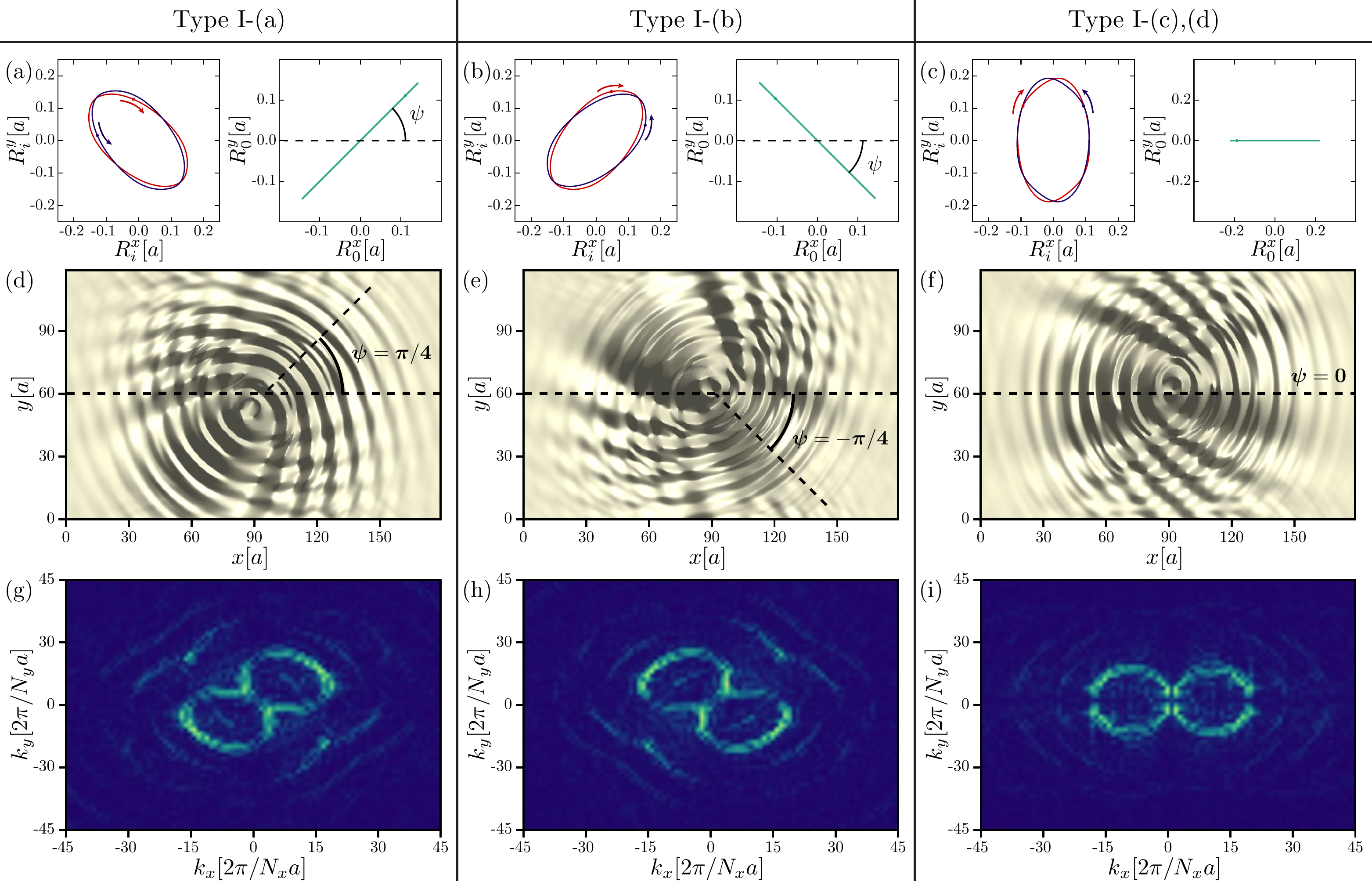}
\caption{Relation between the dynamics of the collective coordinate of position $\mb{R}_i(t)$ and the preferred directionality of the radiated SWs for Type I bilayer that corresponds to the topological charge dipole. In all cases we use $J/D=1$, $b=0.6$ ($B=0.6 JS/g\mu_B$), and $\omega=0.6$ ($\tilde{\omega}=0.6 JS/\hbar$). (a)-(c) Blue lines denote the path of $\mb{R}_i$ for the texture in layer 1, red lines for the one in layer 2, while arrows indicate the sense of gyration. The green lines denote the path of the charge separation distance $\mb{R}_0 = \mb{R}_1 - \mb{R}_2$. (d)-(f) Snapshots of the perpendicular magnetization deviation $\delta n^z_{\mb{r}}(t)$, and (g)-(i) the radiation pattern $\delta n^z_{\mb{k}}$. SWs are radiated by a spin-wave antenna with a preferred direction with respect to $x$ axis, $\psi=\tan^{-1}[\sin(\tilde{\gamma})]$, determined by the helicity of the emitter $\tilde{\gamma}=\gamma_1-\gamma_2$. We find $\psi=\pi/4$ for Type I-(a), $\psi=-\pi/4$ for Type I-(b), and $\psi=0$ for Type I-(c), (d).}
\label{fig:Type_Direction}
\end{figure*}

To simplify the description of the time evolved magnetization, we focus on two salient quantities; the particle's center of mass $\mb{R}_i$ and the amplitude of the far-field emitted SWs. The former corresponds to the collective coordinates of position of the skyrmion core at each layer $i=1,2$ \cite{PAPANICOLAOU1991425},
\begin{align}
R_i^{\nu} = \frac{1}{Q_i} \int d\mb{r}~r^{\nu}~\mb{n}_i \cdot (\pt_x \mb{n}_i \times \pt_y \mb{n}_i)\,,
\label{eq:CenOfMass}
\end{align}
written here for the continuum model, with $\nu=x,y$, while an expression for the discrete model is given in Eq.~\eqref{eq:CollCoorDis}. The emitted SWs correspond to fluctuations of the $z$-component of the magnetization above the oscillating ferromagnetic background, $\delta n^z_{\mb{r}}(t) = n^z_{\mb{r}}(t) - \Mfm(t)$. Unless explicitly stated, we study the fluctuations of the $z$-component $n^{z,1}_{\mb{r}}$ of layer 1, while $n^{z,2}_{\mb{r}}= n^{z,1}_{\mb{r}}$ for Type II and III, and $n^{z,2}_{\mb{r}}=n^{z,1}_{-\mb{r}}$ for Type I. Thus, the excited SWs represent collective modes of the bilayer. Snapshots of $\delta n^z_{\mb{r}}(t)$ for all Types of bilayers are depicted in Fig.~\ref{fig:SW_Types}-(d)-(f), while the full time evolution of the SW pattern is visualized in Supplementary Movies 1, 4, and 5 for Types I-(a), II-(a), and III-(a) respectively. Our simulations clearly show that, collective SW modes are generated by the gyrating interacting topological charges, and propagate from the source to the edge of the sample with characteristics related to the bilayer Type. Type I emits directional, Type II symmetrical, and Type III spiral SWs. 
 
Fig.~\ref{fig:SW_Types} summarizes the relation between the dynamics of the collective coordinate of position $\mb{R}_i(t)$ and the features of the radiated SWs, for all three Types, for $\Jint=0.3 J$, $b=0.6$ ($B=0.6 J S /g\mu_B$), and $\omega=0.6$ ($\tilde{\omega}=0.6 JS/\hbar$). The choice $\omega=b$ signals a resonance condition describing the activation of the CCW and CW mode. For Type I we note that, both the skyrmion (blue line) and the antiskyrmion (red line) perform elliptical paths, with opposite sense of gyration, while the charge separation distance $\mb{R}_0 = \mb{R}_1 - \mb{R}_2$, pointing from the negative to the positive charge, oscillates back and forth along a straight line (green line). This is understood by employing Thiele's approach \cite{PhysRevLett.30.230} to obtain the equation of motion of $\mb{R}_0$ in the limit $Q \gg \alpha$. We then find $-4 \pi Q \epsilon_{\nu \mu} \dot{R}_0^{\mu} =f_0^{\nu}(t)$, where $i=1,2$ is the layer index, $Q$ the charge of layer 1, $\mu,\nu = x,y$, and $\epsilon_{\nu \mu}$ is the antisymmetric tensor. Here $f_0^{\nu}(t)$  parametrizes the relation between $\mb{R}_0$ and the gyrotropic mode activated by the in-plane ac magnetic field (see Supplementary Note \hyperref[sec:Continuum]{3} for a discussion on the Thiele equation).  In view of the numerical results, we use the ansatz $f_0^{\nu}(t) = F_{\nu}(\tilde{\gamma}) \cos(\omega t)$, with $F_x(\tilde{\gamma})= c \sin(\tilde{\gamma}/2)$, and $F_y(\tilde{\gamma})= c \cos(\tilde{\gamma}/2)$, allowing for a dependence on the helicity difference $\tilde{\gamma} = \gamma_1 - \gamma_2$. The solution of the coupled equation of motion equals $\delta R = R_0^y (t) / R_0^x(t) = - F_x/F_y$. For the SB-AN composite pair [Type I-(a)] with $\tilde{\gamma} =-3\pi/2$ we find $\delta R =1$ [see Fig.~\ref{fig:Type_Direction}-(a)], for the SN-AB composite pair [Type I-(b)] with $\tilde{\gamma} =3\pi/2$ we find $\delta R =-1$ [see Fig.~\ref{fig:Type_Direction}-(b)], and for the SN-AN (SB-AB) pair with $\tilde{\gamma} =0$ we find $\delta R = 0 $ [see Fig.~\ref{fig:Type_Direction}-(c)]. Our results suggest that in Type I bilayers, the charge separation distance performs a time-periodic motion on a straight path obtained by rotating the $x$ axis by an angle $\psi=\tan^{-1}[\sin(\tilde{\gamma})]$. 

The second prominent feature when examing Fig.~\ref{fig:SW_Types} is the directionality of the radiated SWs, depending on the type of bilayer system. The behavior of Fig.~\ref{fig:SW_Types}-(d), where we depict a snapshot of $\delta n^z_{\mb{r}}(t)$ for the Type I-(a) bilayer, implies that a topological charge dipole creates directional spin waves, explored further in Fig.~\ref{fig:Type_Direction} for the various subtypes. The preferred radiation direction coincides with the direction of the charge separation path, obtained by rotating the $x$ axis by an angle $\psi=\tan^{-1}[\sin(\tilde{\gamma})]$. This is further supported by the radiation pattern $\delta n^z_{\mb{k}}$ illustrated in Fig.~\ref{fig:Type_Direction}-(g)-(i) [see \hyperref[Meth:RadPat]{Methods} for a definition of $\delta n^z_{\mb{k}}$]. Simple inspection reveals that $\delta n^z_{\mb{k}}$ presents the characteristic two source dipole feature, suggesting that $\delta n^z_{\mb{r}}(t)$ has an azimuthal $\phi$ distribution of the form $\sim \sin(\phi-\psi)$ \cite{Maci2014,Maci__2011,PhysRevB.98.174516}. The helicity $\tilde \gamma$ of the charge dipole can be used to manipulate the propagation direction of the spin waves in the 2D plane. A Type I-(a) bilayer emits waves in the $\psi=\pi/4$ direction, a Type I-(b) in the $\psi=-\pi/4$, and the Types I-(c) and (d) in the $\psi=0$ direction, as illustrated in Fig.~\ref{fig:Type_Direction} and Supplementary Movies 1, 2, and 3, respectively. 
\begin{figure}[t]
\centering
\includegraphics[width=1\columnwidth]{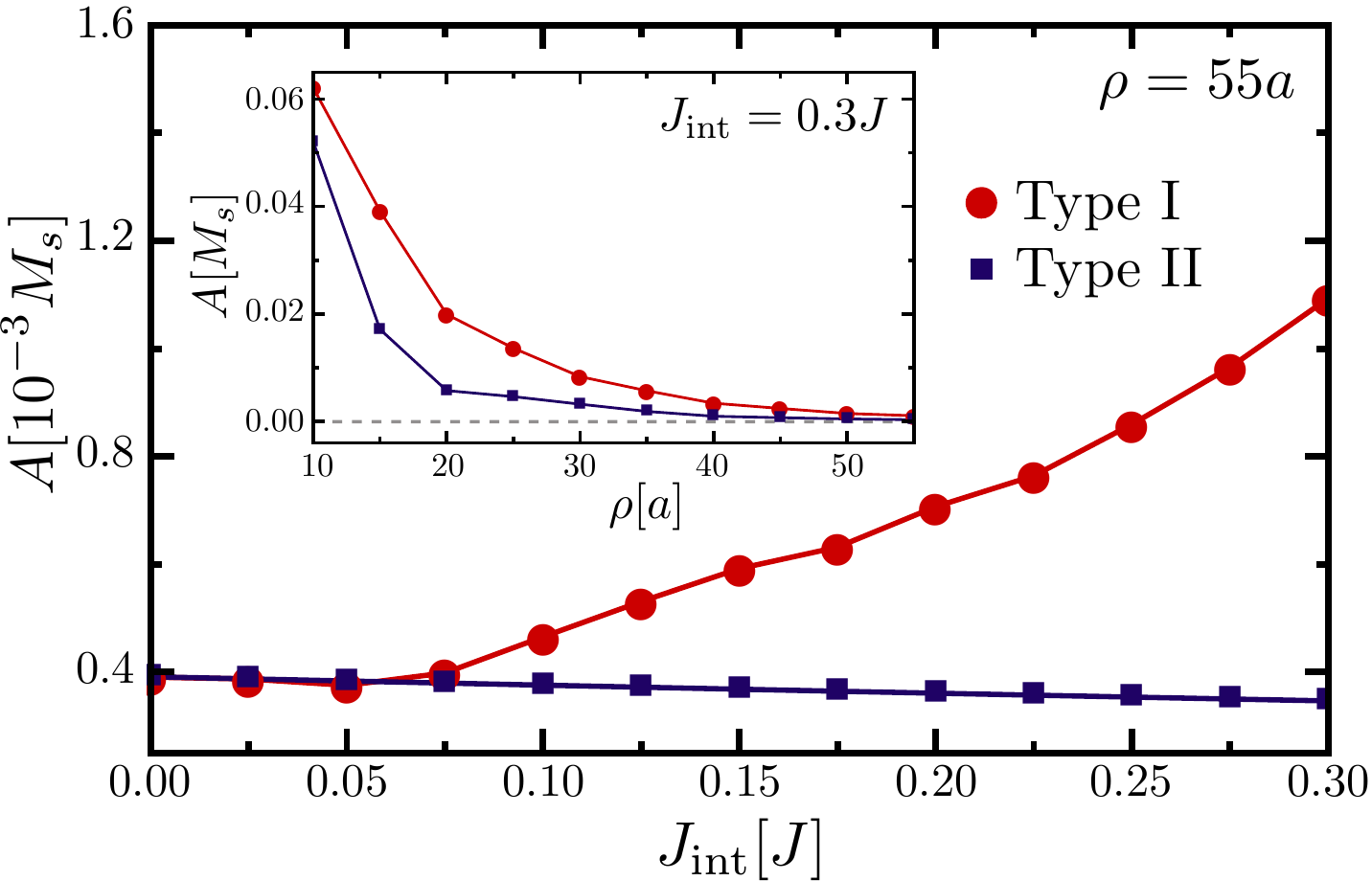}
\caption{Far-field amplitude $A$ as a function of interlayer coupling $J_{\mbox{\scriptsize int}}$ for $\rho=55a$, $b=0.6$ ($B=0.6JS/  g\mu_B$), and $\omega=0.6$ ($\tilde{\omega}=0.6 JS/\hbar$). $A$ increases linearly with $J_{\mbox{\scriptsize int}}$ for Type I bilayers (red line), indicating that the field is radiated from an interacting charge dipole. Type II (blue line) radiation shows a weak dependence on $J_{\mbox{\scriptsize int}}$, and is thus emitted by single monopole sources. The inset depicts the dependence of $A$ on $\rho$ for $J_{\mbox{\scriptsize int}}=0.3 J$, and as expected vanishes fast due to the finite Gilbert damping $\alpha$.}
\label{fig:SWA_Jint}
\end{figure}

In Fig.~\ref{fig:SW_Types}, further results are provided for the SW generation process in Type II and III. In Type II, both textures with the same $Q$ and $\gamma$ perform identical paths, with a vanishing charge separation distance $\mb{R}_0 = 0$. SWs exhibit symmetric radiation patterns [see Fig.~\ref{fig:SW_Types}-(e) and (h)], as the result of a single monopole emitter. Turning now to Type III bilayers, consisting of textures with the same $Q$ and different $\gamma$, although the particles have the same gyrotropic mode, their path $R_i$ depends on $\gamma$ [see Fig.~\ref{fig:SW_Types}-(c)]. A N\'{e}el texture performs a larger circular path (red line), compared to a Bloch texture (blue line), while their charge separation distance $\mb{R}_0$ follows a circular path (green line) with a sense of rotation determined by $Q$. A fascinating feature is the creation of spiral outwards travelling SWs [see Fig.~\ref{fig:SW_Types}-(f) for Type III-(a)].  The physics of the spiral-shape formation is understood in terms of the gyrating motion of $\mb{R}_0$, which plays the role of the source for magnetization oscillations, similarly to other rotating sources that radiate waves with spiral profiles \cite{kapral_showalter_1995,Giordano2016,doi:10.1063/1.3276277}. The direction of the spiral rotation, as illustrated in Fig.~\ref{fig:Spiral} of Supplementary Note \hyperref[sec:Characteristics]{1}, depends on the topological charge $Q$. Spirals of Type III-(a), with $\mb{R}_0$ circulating in a CCW manner, have opposite direction of rotation compared to antispirals of Type III-(b), with $\mb{R}_0$ circulating in a CW manner.
\begin{figure}[t]
\centering
\includegraphics[width=1\columnwidth]{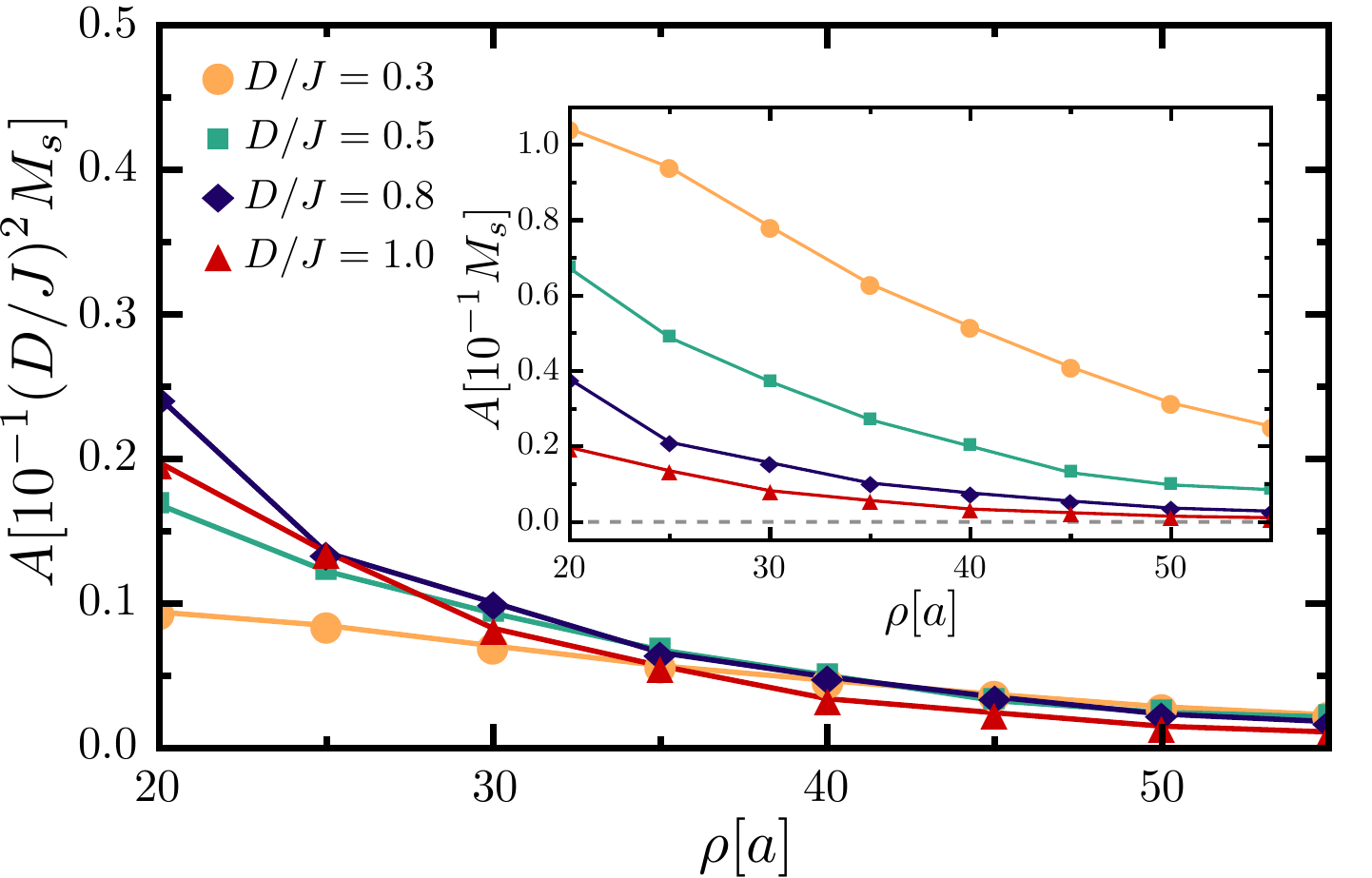}
\caption{Far-field amplitude $A$ of Type I-(a), as a function of $\rho$, for $J_{\mbox{\scriptsize int}}=0.3J $, $b=0.6$ ($B=0.6JS/  g\mu_B$), $\omega=0.6$ ($\tilde{\omega}=0.6 JS/\hbar$), and four values of the ratio $D/J$. $A$ is scaled with $(D/J)^2$ such that the values for different ratios converge for distances sufficiently away from the source, $\rho \gtrsim 30a$.}
\label{fig:SWA_DJ_scaling}
\end{figure}

\begin{figure}[]
\centering
\includegraphics[width=1\columnwidth]{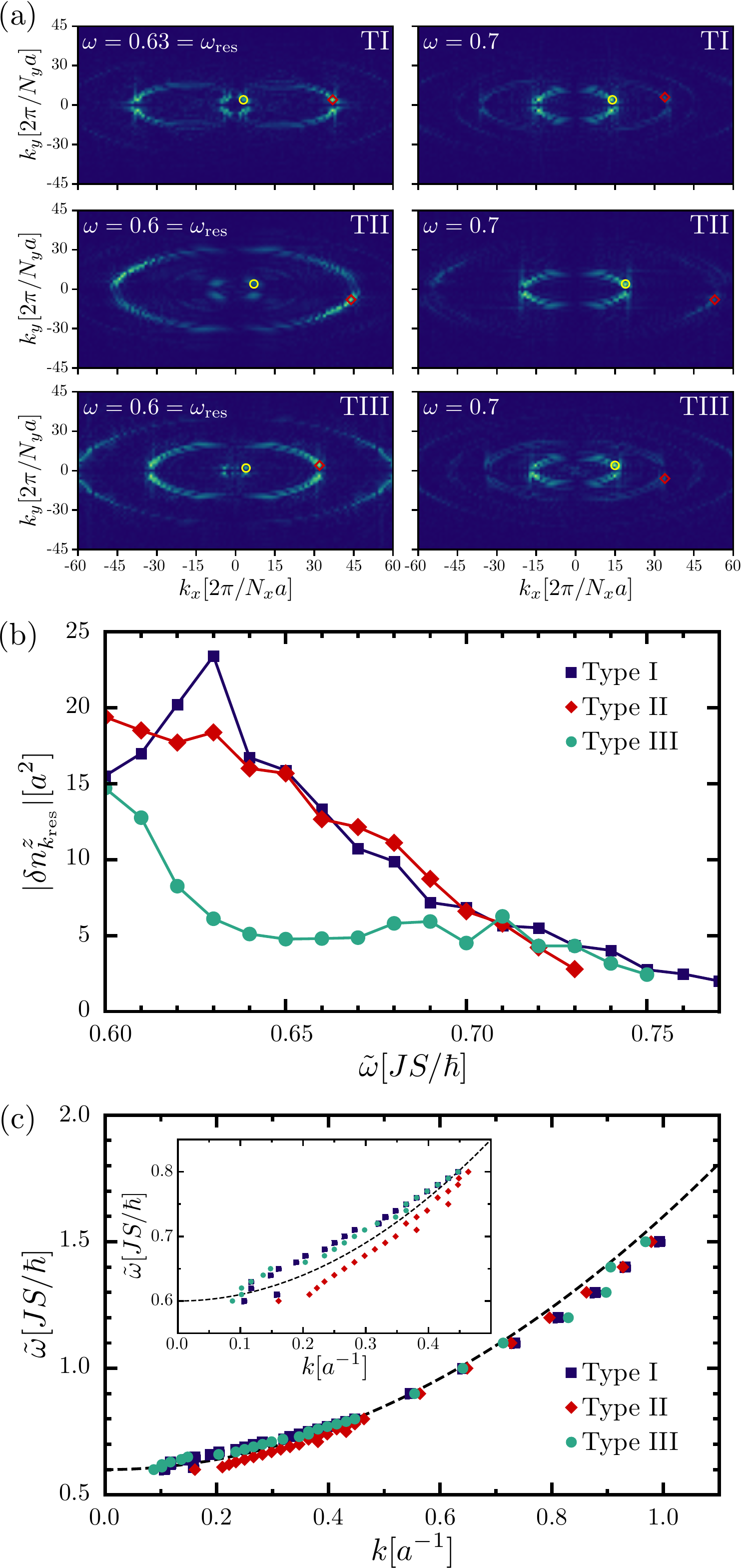}
\caption{(a) Radiation patterns $\delta n_{\mb{k}}^z$ for all Types of bilayers, for $J_{\mbox{\scriptsize int}}=0.3 J$, $b=0.6$, and two frequencies, $\omega=\omega_{\mbox{\scriptsize res}}$ and $\omega=0.7$. The location of the resonant mode $k_{\mbox{\scriptsize res}}$ is indicated by a red rhombus, while the wavevector of the ferromagnetic SW mode is indicated by a yellow circle. (b) The amplitude of the $k_{\mbox{\scriptsize res}}$ mode, $\vert \delta n_{k_{\mbox{\scriptsize res}}}^z \vert$, as a function of the excitation frequency $\omega$. This is a resonant mode with an intensity that vanishes for $\omega>\omega_{\mbox{\scriptsize res}}$. The resonance frequency is $\omega_{\mbox{\scriptsize res}}=0.6$ for Types II and III, and slightly shifted to $\omega_{\mbox{\scriptsize res}}=0.63$ for Type I. (c) Dispersion of the ferromagnetic SWs, for all three Types of bilayers. Solid line indicates the theoretically predicted quadratic relation $\mc{E}/JS = b+a^2k^2$.  The inset depicts small deviations of the recorded energy values from the quadratic dispersion at small $k$ wavevectors.  
}
\label{fig:Identify_Modes}
\end{figure}
~\\
\textbf{Far-Field Spin Wave Amplitude}. We now focus on the characteristics of the on-resonance directional SWs of Type I bilayers, and identify the signatures of the charge interaction in the radiation field. To study the features of the emitted SWs irrespective of a particular direction, we define the far-field SW amplitude as 
\begin{align}
A = \frac{M_s}{N_\mc{J}} \sum_{\mb{r} \in \mc{J}} \frac{1}{2}\left( \max_{ t \in [0,T] }  [ \delta n^z_{\mb{r}}(t)] - \min_{ t \in [0,T] } [ \delta n^z_{\mb{r}}(t)] \right) \,,
\label{eq:SWAmp}
\end{align}
where magnetization is examined over the area $\mc{J}$ of an annulus centered around the skyrmion core, with inner (outer) radius $\rho$ ($P$), comprised of $N_\mc{J}$ lattice sites. $T=2\pi/\omega$ is the period of the ac field, and we use $P=\rho+5 a$ throughout. To capture the characteristics of the SWs generated far from the core, we focus on $\rho \geq 30a$. In the limit $\rho \rightarrow 55a$, thus at the edge of the simulated area, $A \rightarrow 0$ due to the finite Gilbert damping $\alpha$. 

Fig.~\ref{fig:SWA_Jint} summarizes the dependence of the far-field amplitude $A$ on the interlayer coupling $\Jint$ for the Type I (II) bilayer illustrated by the red (blue) line, under a choice of $J/D=1$, $b=0.6$ ($B=0.6 J S/g \mu_B$), and $\rho=55a$. An anticipated result is illustrated, namely the linear dependence of $A$ on $\Jint$ for the Type I bilayer, indicating that the source of the directional SW radiation is indeed an interacting charge dipole, and that changing $\Jint$ is an excellent mechanism to tune the SW amplitude. The inset of Fig.~\ref{fig:SWA_Jint} shows the dependence of $A$ on $\rho$ for $\Jint=0.3 J$, and as expected, for both types of bilayers, vanishes fast due to the finite $\alpha$. For Type II, $A$ shows a weak dependence on $\Jint$, suggesting that even when the layers interact, the SWs remain unaffected by this coupling. Finally, Type III shows an irregular behavior on $\Jint$ presented in Supplementary Note \hyperref[sec:Characteristics]{1}, that could be the result of interference patterns, or that an annulus area $\mc{J}$ is an unsuitable choice for the analysis of emitted SWs with a spiral structure. 

The far-field amplitude $A$ can be enhanced by tuning the $D/J$ ratio, presented in Fig.~\ref{fig:SWA_DJ_scaling}, for four different values of $D/J$. For a fixed value of $\Jint$ and $b$, the skyrmion and antiskyrmion profiles get deformed as the ratio $D/J$ is decreased, in the direction parallel to the axis of SW emission. To avoid deformations, we tune the parameter $\Jint$ to realize textures of approximately the same size. We use $\Jint/J=\{0.3,0.2,0.05,0.012\},$ for $D/J=\{1,0.8,0.5,0.3\}$. It becomes apparent that as $D/J$ is decreased and the system becomes less rigid, the far-field SW amplitude is increased in value, with a $(J/D)^2$ scaling, and vanishes at distances further away from the source. Finally, we reveal an additional mechanism to tune $A$ by varying the external out-of-plane magnetic field $b$. The overall picture suggested by Fig.~\ref{fig:SWA_Bz} given in Supplementary Note \hyperref[sec:Characteristics]{1}, where we plot $A$ as a function of $b$ for $J/D=1$, $\Jint=0.3J$, and $\rho=55a$, is that $A$ is decreased fast by increasing $b$ for both Type I and II. For large magnetic fields $b\gtrsim 0.7$ the spins tend to align along the direction of the field, and local perturbations of the $z$-component vanish. We expect that $A$ can be further manipulated by tuning the layer thickness \cite{PhysRevB.100.214437} and the perpendicular magnetic anisotropy \cite{Zhang2016}, an investigation that we leave for future work. 


~\\
\textbf{Dispersion Characterization}. An estimate of the dispersion relation and the wavelength $\lambda$ of the emitted SWs is obtained by analyzing the radiation patterns $\delta n^z_{\mb{k}}$ for the various bilayer Types as a function of the excitation frequency $\omega$. To minimize finite size effects we consider a larger lattice of $N_x \times Ny$ sites in the $xy$ plane, with $N_x=360$ and $N_y=120$. We examine the results of Fig~\ref{fig:Identify_Modes}-(a), where we plot the radiation pattern $\delta n^z_{\mb{k}}$, for $\Jint=0.3J$, $b=0.6$ ($B=0.6 J S/g\mu_B$), and two frequencies, the resonance frequency $\omega=\wres$, and $\omega=0.7$. We first note that for $\omega=\wres$ there is a distinct short wavelength propagating mode at $\kres=0.65 a^{-1}$ for Type I, $\kres=0.8 a^{-1}$ for Type II, and $\kres=0.57 a^{-1}$ for Type III, with $k^2=k_x^2+k_y^2$ the radial momentum. The location of $\kres$ on the colored surface of $\delta n^z_{\mb{k}}$ of Fig.~\ref{fig:Identify_Modes}-(a) is indicated with a red rhombus, with a corresponding intensity $\delta n^z_{\kres}$. We observe that $\delta n^z_{\kres}$, plotted in Fig.~\ref{fig:Identify_Modes}-(b), takes its maximum at $\wres$ and remains finite for small window of frequencies around $\wres$, before it diminishes for $\omega\gtrsim 0.7$. A small shift on the resonance frequency is observed for Type I, $\wres=0.63$, probably caused by a shift of the CCW (CW) energy at finite $\Jint$. We conclude that $\kres$ is a short wavelength resonant mode activated by the gyrotropic motion of the particle core, further enhanced by the bilayer coupling and with properties linked to the topology of the emitter. 

Besides the resonant mode, we identify a long wavelength mode, with a $k$ vector indicated by a yellow circle in Fig.~\ref{fig:Identify_Modes}-(a). This is the usual SW mode expected for the ferromagnetic layer, with a quadratic dispersion $\mc{E}/JS= b+ a^2k^2$, plotted in Fig.~\ref{fig:Identify_Modes}-(c). We note that, for frequencies away from resonance, the amplitude of this mode dominates over the magnon spectrum, while its wavelength can be tuned by varying the external frequency $\omega$. To reach however, the wavelength of the resonant mode at $\kres=0.63 a^{-1}$ and $\omega=0.6$, a larger excitation frequency $\omega=1$ is required. In the inset of Fig.~\ref{fig:Identify_Modes}-(c), we depict small deviations of the recorded energy values $\mc{E}$ from the quadratic dispersion at small $k$ wavevectors, for all bilayer Types. Whether such deviations are the result of the finite system size considered here, deserves further future investigation. In physical units of $J=1$ meV, $J/D=4$, $\Jint/J=0.3$, $B=324$ mT, $a=1$ nm, $S=1$ and excitation frequency $\omega=57$ GHz, a skyrmion of radius $44$ nm emits resonance modes of wavelength $\lambda = 2\pi J a/D\kres = 42$ nm. Due to the nanoscale of the skyrmion core, the emitted SWs have sufficiently short wavelengths in the sub-100 nm regime. 
 

\section{Discussion}
We have theoretically studied the formation of topological charge dipoles in skyrmion-antiskyrmion bilayers, with a directional SW radiation that exhibits clear dipole signatures in the radiation pattern. The SW emitter is the gyrotropic motion of the interacting skyrmion-antiskyrmion core, activated by ac in-plane magnetic fields. The topological charge separation performs a time-periodic motion along the direction of the SW radiation, determined by the helicity of the source. The resonant magnetic dipole fields are collective modes of the bilayer with sufficiently short wavelengths in the sub-100 nm regime, and a far-field amplitude controlled, among other mechanisms, by the interlayer coupling. SWs with a spiral or antispiral pattern can emerge in bilayers hosting particles with the same $Q$ but different $\gamma$. The origin of the observed SW generation process is attributed to the gyrating motion of the charge separation distance, with a sense of rotation that depends on $Q$. Bilayers with same $Q$ and $\gamma$ form a monopole source and emit radially symmetric waves.

Efficient spin wave emission has been the topic of intensive theoretical and experimental investigations, due to their potential applications in the field of spintronic devices. Arrays of spin torque nano-oscillators (STNO), nanoscale electrical contacts to a ferromagnetic metallic film \cite{Demidov2010}, can be used to create directional spin wave radiation \cite{Maci2014}. In this setup, SW radiation is not symmetrical as the result of interference patterns of excitations originating from two or more STNO placed in an array. In addition, theoretical studies predict that spin waves with a spiral profile are emitted from a gyrotropic rotation of a dynamical skyrmion in metalic Spin-Hall oscillators (SHO) systems \cite{Giordano2016}, in the limit of sufficiently large magnitudes of $D$ and applied charge current. Spiralling spin-wave emission patterns have been experimentally observed in a stack of dynamically excited vortex cores with opposite circulations and parallel cores \cite{Wintz2016,Behncke2018}. 

In our proposed system, waves are radiated along a preferred axis in the 2D plane from a combination of two single emitters in different layers, forming a coupled bound state that corresponds to a topological charge dipole. Such a mechanism does not rely on spin wave interference, and is thus more reliable and robust. It also offers more advantages, since it can be realized in a large variety of skyrmion-hosting materials, including magnetic insulators, relevant for high-power applications. It is also independent of mechanisms that require particular fabrication of magnetic elements with tailored properties, as is done in magnonic crystals \cite{PhysRevB.86.144402,doi:10.1063/1.4737438}. Besides the fundamental interest of our results, we anticipate our findings to lead to the development of novel efficient SW emitters and SW antennas, with tunable characteristics linked to the topology of the source.


\section{Methods} 

\subsection{Micromagnetic Simulations} \label{Meth:MicroMagn}

To simulate the magnetization dynamics of the insulating bilayer, we numerically solve the Landau-Lifshitz-Gilbert (LLG) equation, 
\begin{align}
\frac{d \mb{n}^i_\mb{r}}{dt} = -\frac{\gamma_0 \mb{n}^i_\mb{r}}{1+\alpha^2}\times\left( \mb{H}_{\scalebox{0.7}{eff}} + \alpha \mb{n}^i_\mb{r}\times  \mb{H}_{\scalebox{0.7}{eff}}  \right)\,,
\label{eq:LLG}
\end{align}
with $\mb{H}_{\scalebox{0.7}{eff}}=(1/\gamma_0 \hbar)\pt H/\pt  \mb{n}^i_\mb{r} $, for each of the $i=1,2$ layers. We consider two coupled monolayers of a finite lattice of $180 \times 120$ sites in the $xy$ plane. Here $\mb{n}^i_\mb{r}=\mb{S}^i_\mb{r}/M_s$ with $M_s = g\mu_B S/a^2 d$ the saturation magnetization, $\gamma_0$ is the gyromagnetic ratio, $\alpha$ is the Gilbert damping coefficient describing spin relaxation, and $d$ the layer thickness. Time $t$, frequency $\omega$, and space $\mb{r}$ are given in dimensionless units. Physical units are restored as $\tilde{t}= \hbar t /JS$, $\tilde{\mb{r}}=\mb{r} a$, and $\tilde{\omega}=JS \omega/\hbar$. $H$ is the total Hamiltonian including an external time-oscillating magnetic field along the $x$ axis, $H = H_T + \Hosc$, with $H_T= H_1+H_2 +\Hint$ and $\Hosc= -\sum_{\mb{r}} g\mu_B B_0 \cos(\omega t) (n_{\mb{r}}^{x,1}+n_{\mb{r}}^{x,2})$. The simulations were performed using $J=1$, $\alpha=0.08$, $b_0 = g\mu_B B_0/JS=0.1$, and unless explicitly stated, $D=1$. The value of the uniform field ranges between $0.6 \leq b \leq 0.9$, of the interlayer coupling between $0\leq \Jint\leq0.3$, of the external frequency between $0.6\leq \omega \leq 0.85$. The initial spin configurations supported by the Hamiltonian Eq.~\eqref{eq:SpinH}, before the onset of the oscillating field, correspond to spin textures carrying a finite $Q$ and are calculated by means of a Monte Carlo simulated annealing method. They correspond to initial states for the simulation of the time-evolution of magnetic moments of Eq.~\ref{eq:LLG}. We numerically verify that a time-oscillating field along the $x$ direction activates the CCW mode for a skyrmion and the CW mode for an antiskyrmion, signaled by a resonance peak at $\omega \simeq b$. 

\subsection{Skyrmion/Antiskyrmion Center of Mass} \label{Meth:CenterofMass}

The collective coordinate of position $R_i^{\nu}$, for a spin field $\mb{n}_{\mb{r}}^i$ defined on a discrete square lattice, is given by 
\begin{align} \label{eq:CollCoorDis}
R_i^{\nu} = \frac{1}{Q_i} \sum_\mb{r} (\mb{r}\cdot\mb{e}_{\nu} + \tfrac{1}{2}a ) \sigma^i_\mb{r} \,.
\end{align}
Here
\begin{align}
\sigma^i_\mb{r} = (\Omega^i_{\mb{r},\mb{r} + a\mb{e}_{x},\mb{r} + a\mb{e}_{x} + a\mb{e}_{y}} + \Omega^i_{\mb{r},\mb{r} + a\mb{e}_{x} + a\mb{e}_{y},\mb{r} + a\mb{e}_{y}})/4\pi
\end{align}
is the discretized topological charge \cite{VanOosterom1983}  over a square plaquette centered at $\mb{r} + \tfrac{1}{2}a(\mb{e}_{x} + \mb{e}_{y})$, computed in terms of the solid angles subtended by the spins at sites $\{ \mb{r},\mb{r} + a\mb{e}_{x},\mb{r} + a\mb{e}_{x} + a\mb{e}_{y} \}$ and $\{ \mb{r},\mb{r} + a\mb{e}_{x} + a\mb{e}_{y},\mb{r} + a\mb{e}_{y} \}$ using the expression \cite{Berg1981}
\begin{align}
\tan \big[ \half \Omega^i_{\mb{r}_1,\mb{r}_2,\mb{r}_3} \big] = \frac{ \mb{n}^i_{\mb{r}_1} \cdot \mb{n}^i_{\mb{r}_2} \times \mb{n}^i_{\mb{r}_3} }{ 1 + \mb{n}^i_{\mb{r}_1} \cdot \mb{n}^i_{\mb{r}_2} + \mb{n}^i_{\mb{r}_1} \cdot \mb{n}^i_{\mb{r}_3} + \mb{n}^i_{\mb{r}_2} \cdot \mb{n}^i_{\mb{r}_3}  } \,.
\end{align}
Finally, $Q_i = \sum_\mb{r} \sigma^i_\mb{r}$ is the topological charge of $\mb{n}_{\mb{r}}^i$ over the entire lattice. 

\subsection{Fourier Transform Radiation Pattern} \label{Meth:RadPat}

In order to reveal the radiation pattern of the emitted spin waves we used $\delta n^z_{\mb{k}}$, the absolute value of the Fourier Transform of $\delta n^z_{\mb{r}}(t) = n^z_{\mb{r}}(t) - \Mfm(t)$, averaged over one period of the ac driving field. To further enhance the far-field spin wave amplitude features, the sign of the Hilbert Transform of $\delta n^z_{\mb{r}}(t)$ was obtained before taking the Fourier Transform. The full expression is given below
\begin{align}
\delta n^z_{\mb{k}} = \frac{1}{T} \int_0^T dt \left| \mathcal{F}_\mb{k}[ \mbox{sign}(\mathcal{H} [ \delta n^z_{\mb{r}}(t) ])] \right|\,. 
\end{align}

\section{Acknowledgments}
T.H. is grateful to T. Hinokihara for useful discussions. T.H. was supported by the Japan Society for the Promotion of Science through Program for Leading Graduate Schools (MERIT), JSPS KAKENHI (Grant No. 16J07110) and Young Researchers' Exchange Program between Japan and Switzerland 2018. C.P. has received funding from the European Union's Horizon 2020 research and innovation programme under the Marie Sklodowska-Curie grant agreement No 839004. S.A.D. and D.L were supported by the Swiss National Science Foundation (Switzerland) and the NCCR QSIT. 



\bibliography{SWETopoDipole}


\setcounter{figure}{0}
\renewcommand{\thefigure}{S\arabic{figure}}
\renewcommand{\theHfigure}{\thefigure}

\setcounter{equation}{0}
\renewcommand{\theequation}{S\arabic{equation}}
\renewcommand{\theHequation}{\theequation}

\clearpage


\section{Supplementary Note 1: Spin Wave Characteristics}
\label{sec:Characteristics}
\begin{figure}[b!]
\centering
\includegraphics[width=1\columnwidth]{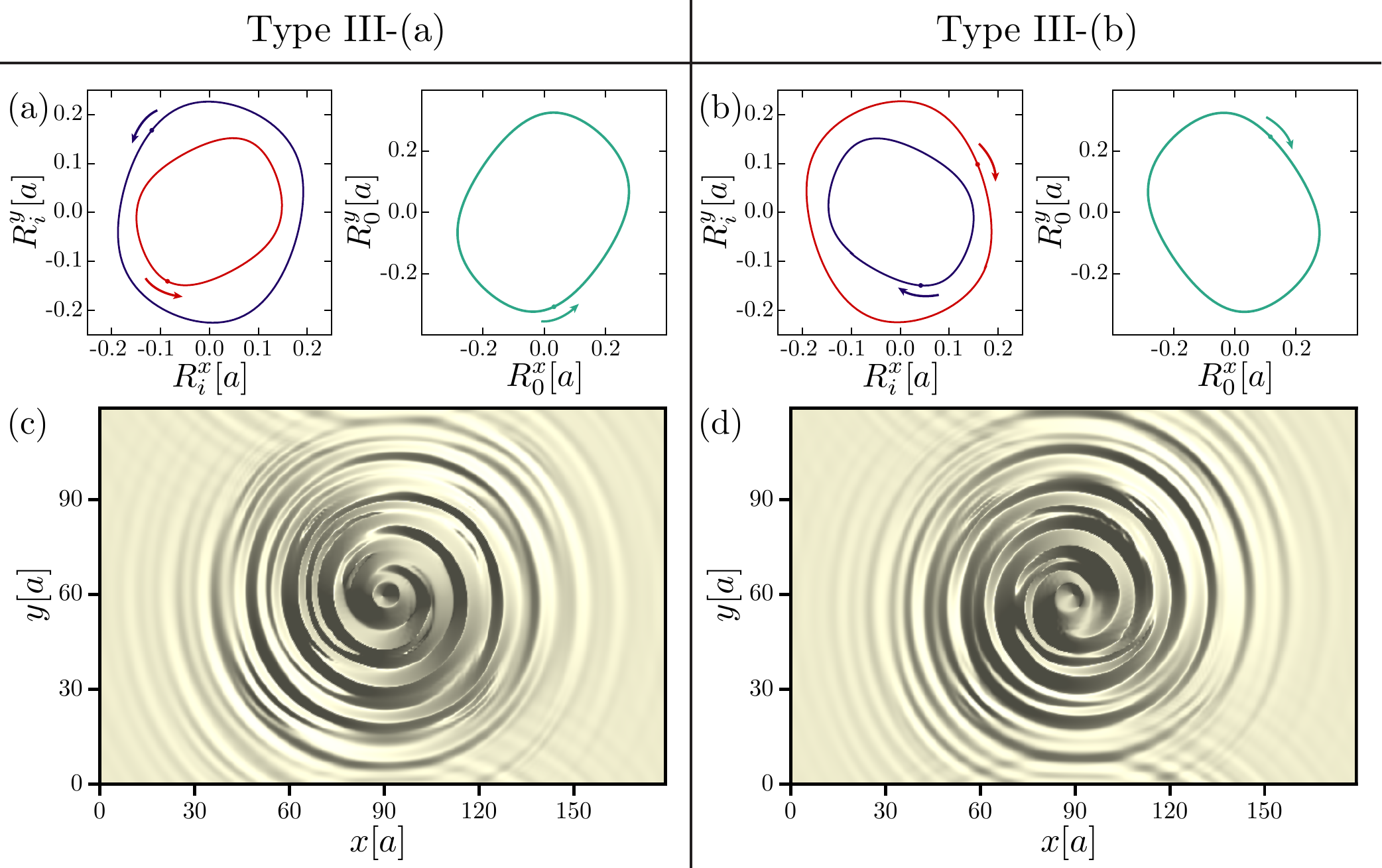}
\caption{Spirals and Antispirals. The direction of spiral rotation depends on the topological charge $Q$. (a) The charge seperation distance $\mb{R}_0$ rotates in a CCW manner in Type III-(a) bilayers, (b) and in a CW manner in Type III-(b) bilayers. (c) The direction of the spiral SW of Type III-(a) is opposite compared to (d) antispirals of of Type III-(b).}
\label{fig:Spiral}
\end{figure}

\begin{figure}[t!]
\centering
\includegraphics[width=1\columnwidth]{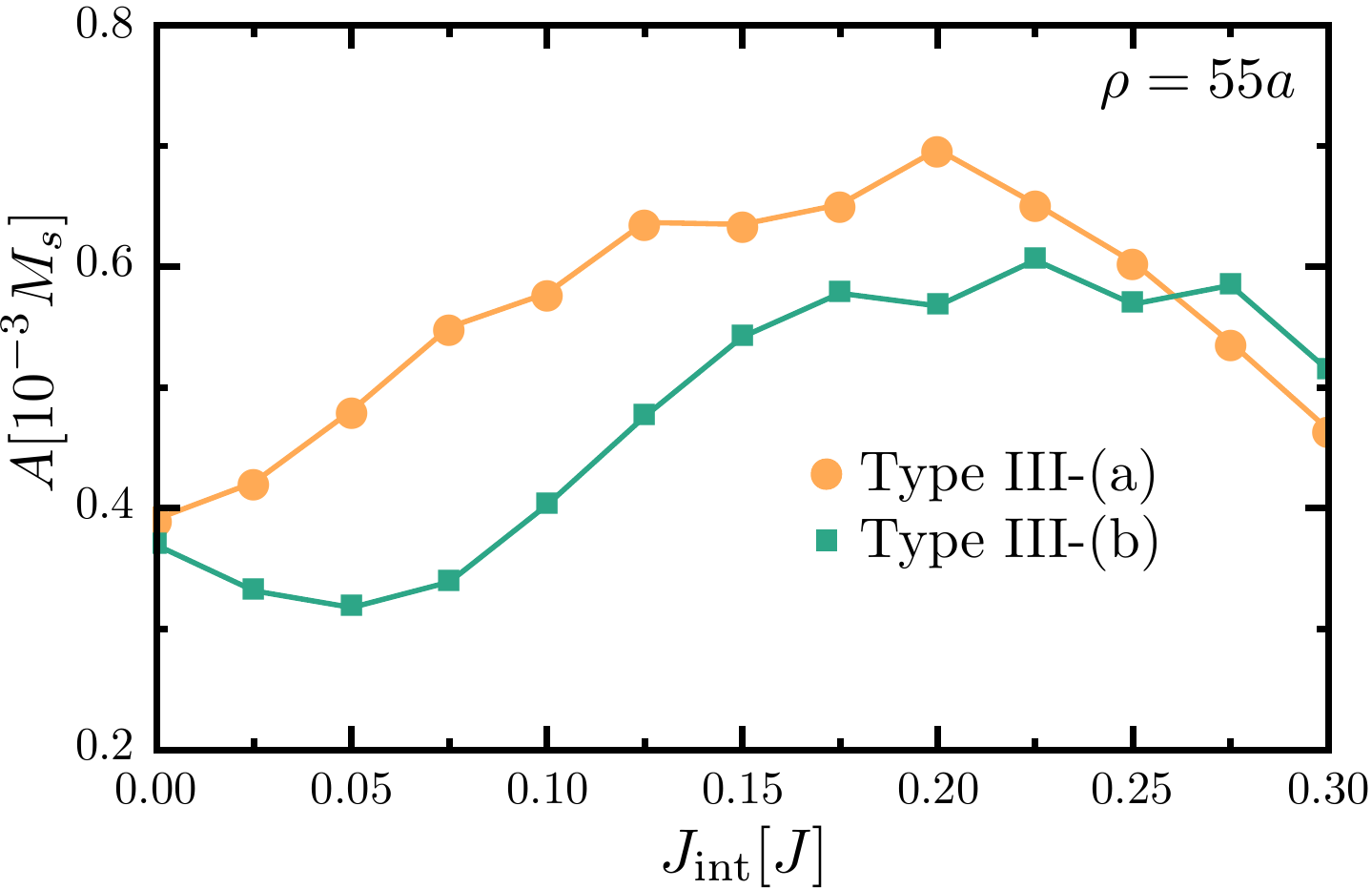}
\caption{Far-field amplitude $A$ as a function of interlayer coupling $J_{\mbox{\scriptsize int}}$ for Type III, $\rho=55a$, and $b=0.6=\omega$. $A$ shows an irregular behavior on $J_{\mbox{\scriptsize int}}$, that could be the result of interference patterns, or an unsuitable choice of an annulus area $\mathcal{J}$ for the analysis of the emitted SWs with a spiral structure.}
\label{fig:SWA_Jint_TypeIII}
\end{figure}

In this Supplementary Note we provide further details on the characteristics of the emitted SWs. For Type III bilayers, SWs have a spiral structure with a preferred sense of rotation determined by the sign of the topological charge of the source, as is illustrated in Fig.~\ref{fig:Spiral}. For two layers hosting skyrmions (antiskyrmions) with $Q=-1$ ($Q=1$), the charge separation distance $\mb{R}_0$, which plays the role of the source, performs a CCW (CW) rotation, and the emitted SWs have a spiral (antispiral) shape. To verify whether the spiral SWs are emitted by a source consisting of two interacting particles, in Fig.~\ref{fig:SWA_Jint_TypeIII} we plot the far-field amplitude $A$ as a function of the interlayer coupling $\Jint$, for $\rho = 55 a$. For Type III-(a), $A$ is an increasing function for $\Jint < 0.2$, while it decreases for $\Jint \geq 0.2$. On the contrary, for Type III-(b) $A$ has a small decrease for small $\Jint$ and later increases at a slow rate. This irregular behavior on $\Jint$ could be the result of interference patterns, or unsuitable choice of an annulus area $\mc{J}$ for the analysis of emitted SWs with a spiral structure. To obtain conclusive results, a systematic way to exclude interference patterns, as well as the consideration of different geometries $\mc{J}$, need to be employed. 

\begin{figure}[t!]
\centering
\includegraphics[width=1\columnwidth]{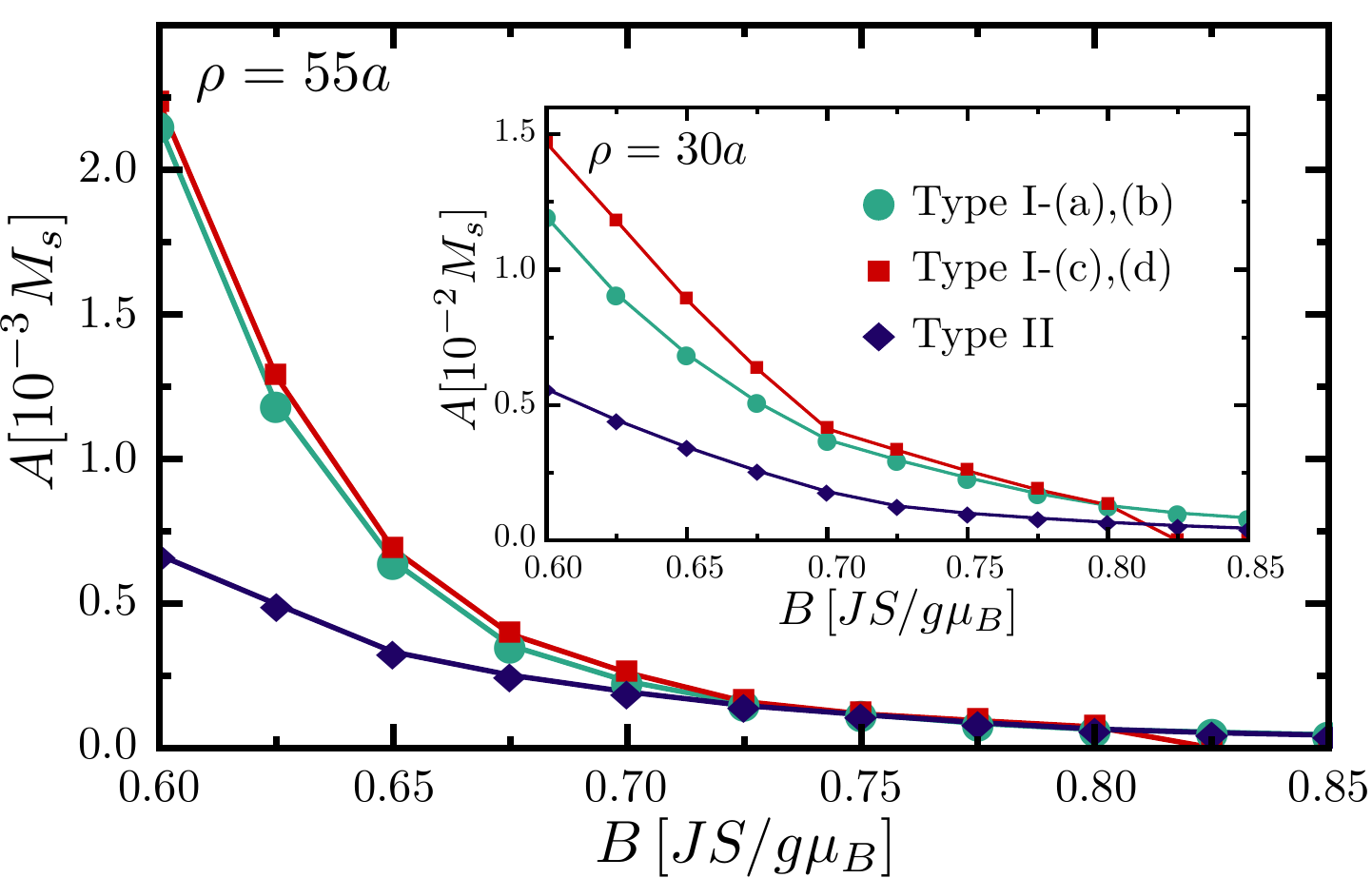}
\caption{Far-field amplitude $A$ as a function of external out-of-plane magnetic field $b$ for $\rho=55a$, $J_{\mbox{\scriptsize int}}=0.3J $, and on-resonance frequency $\omega=b$. $A$ decreases fast with $b$ for both Type I (red line) and II (blue line) bilayers, indicating that for large magnetic fields, local perturbations of the $z$-component vanish. The inset reports a small difference of the value of $A$ between Type I-(a)-(b) (red line) and Type I-(c)-(d) (green line) for $\rho=30 a$ closer to the source.}
\label{fig:SWA_Bz}
\end{figure}
Besides changing the interlayer coupling $\Jint$ and the ratio $J/D$, an additional mechanism to tune the far-field amplitude $A$ is by varying the external out-of-plane magnetic field $b$. In Fig.~\ref{fig:SWA_Bz}, we plot $A$ as a function of $b$ for $J/D=1$, $\Jint=0.3 J$, and $\rho=55a$. As expected, $A$ is decreased fast by increasing $b$ for both Type I and II. For large magnetic fields $b\gtrsim 0.7$ the spins tend to align along the direction of the field, and local perturbations of the $z$-component vanish. We emphasize that no variation in the value of $A$ appears between the various subtypes of Type I and II bilayer. A small difference of the value of $A$ between Type I-(a)-(b) and Type I-(c)-(d) (green line) is reported in the inset of Fig.~\ref{fig:SWA_Bz}, when measured closer to the source for $\rho=30 a$. 


\section{Supplementary Note 2: Spin Wave Hamiltonian}
\label{sec:Magnons}
In the main text we argued that the sense of gyration of localized deformations of the magnetization profile depends on the sign of $Q$, which we now substantiate by a numerical calculation of the magnon spectrum for all types of magnetic textures studied here. We consider the spin-lattice Hamiltonian of Eq.~\eqref{eq:SpinH} with $\{\mb{d}_{\mb{e}_x},\mb{d}_{\mb{e}_y}\}=\{-\mb{e}_x,-\mb{e}_y\}$ for cubic, $\{\mb{d}_{\mb{e}_x},\mb{d}_{\mb{e}_y}\}=\{-\mb{e}_y,\mb{e}_x\}$ for interfacial, $\{\mb{d}_{\mb{e}_x},\mb{d}_{\mb{e}_y}\}=\{\mb{e}_y,\mb{e}_x\}$ for $C_{2v}$, and $\{\mb{d}_{\mb{e}_x},\mb{d}_{\mb{e}_y}\}=\{\mb{e}_x,-\mb{e}_y\}$ for $D_{2d}$ crystal symmetry. Quantum spin fluctuations are described using Holstein-Primakoff (HP) bosons. In the case of noncollinear textures, such as those investigated here, it is necessary to first choose the spin quantization axis along the direction of the classical ground state $\mb{n}_\mb{r}$. This is accomplished by introducing an orthonormal basis at each lattice site, $\{ \mb{m}^1_\mb{r}, \mb{m}^2_\mb{r}, \mb{n}_\mb{r} \}$, with $\mb{m}^1_\mb{r} \times \mb{m}^2_\mb{r} = \mb{n}_\mb{r}$. We can now introduce the rotated spin operators at each site according to $\mb{S}_\mb{r} = \mb{m}^1_\mb{r} \SS_\mb{r}^1 + \mb{m}^2_\mb{r} \SS_\mb{r}^2 + \mb{n}_\mb{r} \SS_\mb{r}^3$. The HP transformation at site $\mb{r}$ then reads
\begin{eqnarray}
\SS_\mb{r}^+ &=& (2S - a_\mb{r}^\dag a_\mb{r})^{1/2} a_\mb{r} \,,\\
\SS_\mb{r}^- &=& a_\mb{r}^\dag(2S - a_\mb{r}^\dag a_\mb{r})^{1/2} \,,\\
\SS_\mb{r}^3 &=& S - a_\mb{r}^\dag a_\mb{r} \,,
\end{eqnarray}
where $\SS_\mb{r}^\pm = \SS_\mb{r}^1 \pm i \SS_\mb{r}^2$, and with the HP boson operators satisfying the bosonic algebra: $[a_\mb{r}, a_{\mb{r}'}^\dag] = \delta_{\mb{r},\mb{r}'}$ and $[a_\mb{r}, a_{\mb{r}'}] = 0 = [a_\mb{r}^\dag, a_{\mb{r}'}^\dag]$. Following a standard procedure, the spin-lattice Hamiltonian is expanded as a series in $1/S$. The spin wave Hamiltonian, identified as the $\CalO (S)$ piece, has the form 
\begin{align}\label{eq:Hsw_reciprocal}
\Hsw = \half S \sum_{\mb{r},\mb{r}'}
\mc{X}_{\mb{r}}^{\dagger} \begin{pmatrix}
\Omega_{\mb{r},\mb{r}'} & - \Delta_{\mb{r},\mb{r}'} \\
- \Delta^*_{\mb{r},\mb{r}'} & \Omega^*_{\mb{r},\mb{r}'}
\end{pmatrix} \mc{X}_{\mb{r}'}
- \half S\sum_\mb{r} \Lambda_\mb{r} \,,
\end{align}
where $\mc{X}_{\mb{r}}^{\dagger} = \big( \, a_{\mb{r}}^\dag \,,\, a_{\mb{r}} \, \big)$, $\Lambda_\mb{r} = \sum_{\mb{r}'} \big[ J_{\mb{r},\mb{r}'} \, (\mb{n}_\mb{r} \cdot \mb{n}_{\mb{r}'}) + \BD_{\mb{r},\mb{r}'} \cdot (\mb{n}_\mb{r} \times \mb{n}_{\mb{r}'}) \big] + \frac{g\mu_B B}{S} n^z_\mb{r}$, $\Omega_{\mb{r},\mb{r}'} =  \delta_{\mb{r},\mb{r}'} \Lambda_\mb{r} - \half \big[ J_{\mb{r},\mb{r}'} (\mb{m}_{\mb{r}}^+ \cdot \mb{m}_{\mb{r}'}^-) + \BD_{\mb{r},\mb{r}'} \cdot (\mb{m}_{\mb{r}}^+ \times \mb{m}_{\mb{r}'}^-) \big]$, $\Delta_{\mb{r},\mb{r}'} = \half \big[ J_{\mb{r},\mb{r}'} (\mb{m}_{\mb{r}}^+ \cdot \mb{m}_{\mb{r}'}^+) + \BD_{\mb{r},\mb{r}'} \cdot (\mb{m}_{\mb{r}}^+ \times \mb{m}_{\mb{r}'}^+) \big]$, with $J_{\mb{r},\mb{r}'} = J(\delta_{\mb{r} - \mb{r}', \pm a\mb{e}_x} + \delta_{\mb{r} - \mb{r}', \pm a\mb{e}_y})$, $\BD_{\mb{r},\mb{r}'} = D(\mb{d}_{\pm \mb{e}_x} \delta_{\mb{r} - \mb{r}', \pm a\mb{e}_x} + \mb{d}_{\pm \mb{e}_y} \delta_{\mb{r} - \mb{r}', \pm a\mb{e}_y})$, and $\mb{m}_{\mb{r}}^\pm = \mb{m}_{\mb{r}}^1 \pm i\mb{m}_{\mb{r}}^2$.

In the following we discuss the spectrum and form of excitations around the equilibrium magnetization profile, which can be either a skyrmion or an antiskyrmion, obtained by a numerical diagonalization of Eq.~\eqref{eq:Hsw_reciprocal}, performed for $J/D=1$ on a square of $30 \times 30$ spins with periodic boundary conditions. 
\begin{figure}[t!]
\centering
\includegraphics[width=\columnwidth]{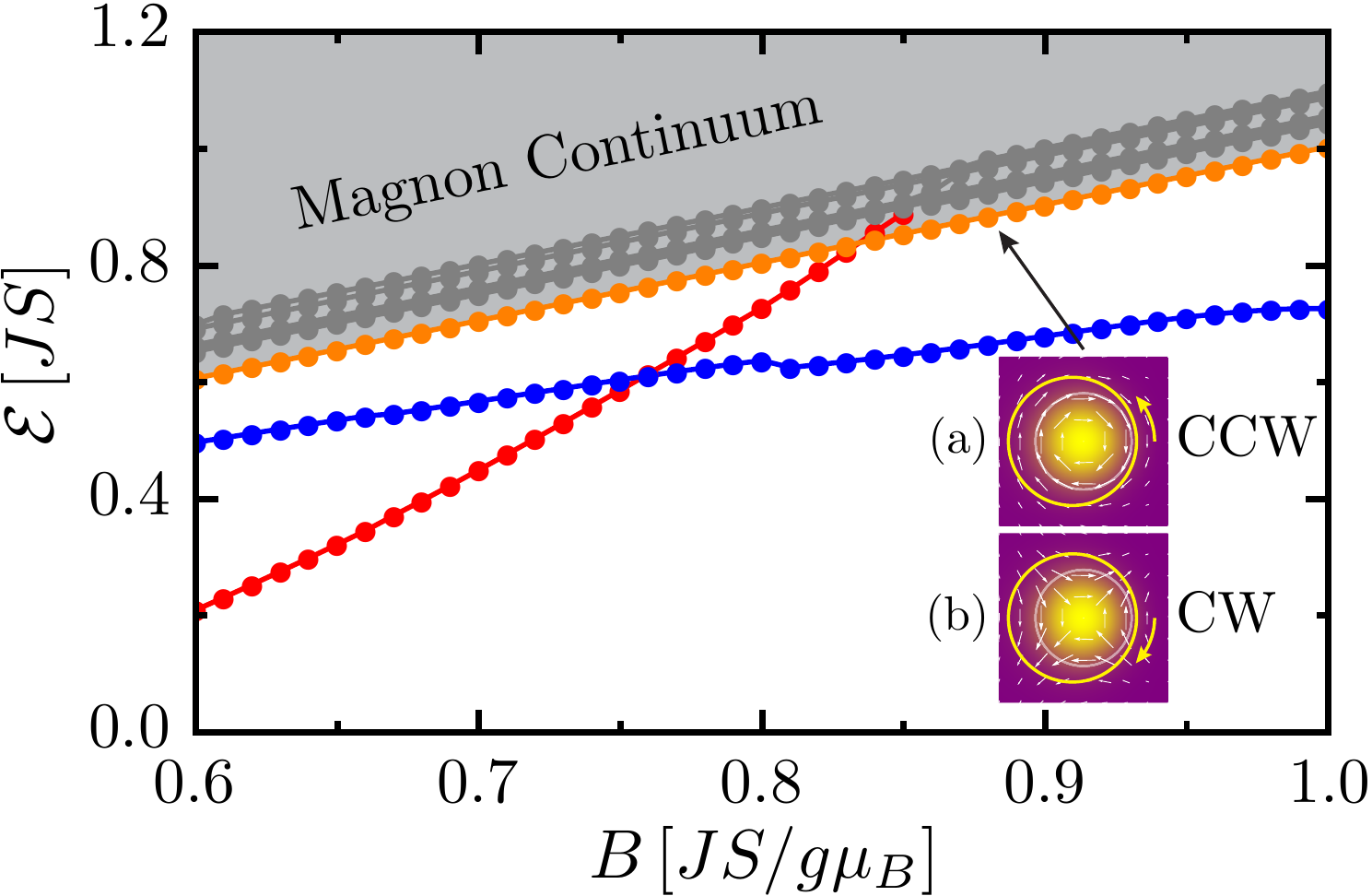}
\caption{Magnetic field dependence of the magnon energy $\mc{E}$, derived by a numerical diagonalization of the spin wave Hamiltonian \eqref{eq:Hsw_reciprocal}, for $J/D=1$, on a square of $30 \times 30$ spins, in the presence of a single skyrmion with $\gamma=\pi/2$ and $Q=-1$. In addition to the magnon continuum (shaded region) with a boundary $\varepsilon_{\mbox{\tiny gap}} =b$, we find a breathing (blue line), an elliptical (red line), and a counter-clockwise (CCW) localized mode (orange line), visualized in inset (a). The magnon energy is independent of the choice of $\gamma$ and $Q$, but the sense of gyration for the corresponding eigenstates depends on the sign of $Q$. Thus, the orange line corresponds a clockwise (CW) mode for an antiskyrmion, visualized in inset (b).  }
\label{fig:MagnonSpec}
\end{figure}
In Fig.~\ref{fig:MagnonSpec}, we depict the energies of the 10 lowest-lying magnon modes as a function of the external magnetic field. First we note that the magnon spectrum is insensitive to the choice of $Q$ and $\gamma$; any combination of topological charge and helicity reproduces the same energy spectrum. We observe however, that the sense of gyration of localized deformations of the magnetization profile depends on the sign of $Q$. Local modes of the skyrmion with a CCW sense of gyration, correspond to CW modes for the antiskyrmion. This observation is confirmed by an analytical derivation of the magnon eigenvalue problem derived in the continuum model, given explicitly in Supplementary Note \hyperref[sec:Continuum]{3}.

Several important facts have become apparent in Fig.~\ref{fig:MagnonSpec}, which we analyze. The grey shaded area represents the magnon continuum which exists above the gap due to the magnetic field $\egap = g\mu_B B$. Below the gap of the scattering states, we find three bound states, two of them correspond to the breathing mode (blue line) and the elliptical (red line). Of particular importance for the present study is the localized state with energy just below the magnon gap (orange line), $\mc{E}_0 \approx \egap$, which corresponds to the counterclockwise (CCW) mode for the skyrmion, and as already suggested, to the clockwise (CW) mode for the antiskyrmion. These modes correspond to a rotation of the out-of-plane spin components around the (anti)skyrmion core. The CCW and CW modes are excited by an in-plane ac magnetic field, while exciting the breathing mode requires an out-of-plane ac magnetic field. Finally, a zero energy mode is found, associated with translations of the skyrmion position in the 2D plane (not shown). 


\section{Supplementary Note 3: Continuum limit}
\label{sec:Continuum}

Here we present the continuum model of the discrete magnetic Hamiltonian of Eq.~\eqref{eq:SpinH}, the corresponding stable solutions and the structure of the magnon eigenvalue problem, in order to demonstrate that the sense of gyration of localized deformations of the magnetization profile depends on the topological charge $Q$. To derive the classical energy functional in the continuum, valid in the limit of slowly varying magnetic textures, we treat the spin operators as classical vectors of length $S$, $\mb{S}_\mb{r} \to S\mb{n}_\mb{r}$, where $\mb{n}_\mb{r}$ is a unit vector. In the limit $a \rightarrow 0$, with $a$ being the lattice spacing, $\mb{r}$ becomes a continuous variable and $\mb{n}_\mb{r}$ turns into a field $\mb{n}(\mb{r})$. The resulting classical magnetic energy is,
\begin{align}
\mc{W}= \int d\mb{r} [\tilde{J} (\nabla \mb{n})^2 + \Edm - B_z n_z] \,,
\label{eq:ContModel}
\end{align}
where $\Edm$ is the Dzyaloshinskii-Moriya (DM) interaction and it is equal to,
\begin{align}
\EC= \mc{D} [\mc{L}_{yz}^{(y)}-\mc{L}_{xz}^{(x)}]\,, \ \  \ED=- \mc{D} [\mc{L}_{xz}^{(y)}+\mc{L}_{yz}^{(x)}] \label{eq:EDMIa} \\
 \EB=-\mc{D} [\mc{L}_{xz}^{(y)}+\mc{L}_{zy}^{(x)}]\,, \quad \EIF=\mc{D} [\mc{L}_{zx}^{(x)}+\mc{L}_{zy}^{(y)}] \label{eq:EDMIb} 
\end{align}
 for the $C_{2v}$ symmetry, the $D_{2d}$ symmetry, the cubic, and the interfacial DM interaction, respectively. The Lifshitz invariant is denoted as $\mc{L}_{ij}^{(k)}= n^{i} \pt_n^{j}/ \pt_x^k- n^{j} \pt_n^{i}/ \pt_x^k$. The parameters in \eqref{eq:ContModel} are related to those in \eqref{eq:SpinH} via $\{ \tilde{J}, \mc{D}, B_z \} = \{ S^2J, S^2D/a, g\muB S B/a^2 \}$, and without loss of generality we assume $\mc{D}>0$. Using the spherical parametrization $\mb{n}=[\sin \Theta \cos \Phi,  \sin \Theta \sin \Phi, \cos\Theta]$, (anti)skyrmions appear as particle-like metastable solutions of the functional in Eq.~\eqref{eq:ContModel}, described by $\Phi(\mb{r})=\mu \phi+ \gamma$ and the approximate function $\Theta (\rho)= 2 \tan^{-1}[(\lambda_0/\rho) e^{-(\rho -\lambda_0)/\rho_0}]$, with $\mb{r}=(\rho,\phi)$ the polar coordinate system, $\rho_0 = \sqrt{2 \tilde{J}/B_z}$, while $\lambda_0$, which we obtain numerically from the Euler-Lagrange equation of the stationary skyrmion, is the skyrmion radius. 

\textit{Magnetic Excitations}. Next we consider fluctuations around the static skyrmion as $\Phi =\Phi_0 + \xi= \mu \phi+\gamma +\xi$ and $\Theta = \Theta_0 + \eta$ and rewrite the energy functional as $\mc{W} = \mc{W}_0 + \chi^{\dagger} \mc{H} \chi$, where $\mc{W}_0(\Phi_0, \Theta_0)$ is the configuration energy functional of the (anti)skyrmion field, and $\mc{H}$ is the magnon Hamiltonian calculated for the convenient spinor representation $\chi =1/2 (\xi \sin \Theta_0- i \eta , \xi \sin \Theta_0+ i \eta)^{T}$. Magnon states are found by solving the eigenvalue problem (EVP) $\mc{H} \Psi_{n} = \mc{E}_n \sigma_z \Psi_n$, while it appears convenient to represent these solutions in terms of wave expansions $\Psi_n = e^{i m \phi} \psi_{n,m}(\rho)$. The EVP is written as $\mc{H}_m  \psi_{n,m}(\rho) = \mc{E}_{n,m} \sigma_z  \psi_{n,m}(\rho)$, with 
\begin{align}
\mc{H}_m = \tilde{J} (- \nabla^2_{\rho}+ U_0(\rho) +\frac{m^2}{\rho^2} )\mathds{1} + W(\rho) \sigma_x  + \mu m V(\rho) \,, 
\label{eq:MagnHam}
\end{align}
where $V(\rho)$, $W(\rho)$ and $U_0(\rho)$ are potentials of the radial coordinate $\rho$ (for explicit expressions see Ref.~\cite{PhysRevB.100.134404}, and in particular Eqs.~(D1)--(D3) therein). Solutions of the EVP include propagating scattering states with eigenfrequencies above the magnon gap $\egap=g \mu_B B$, as well as massive internal modes that are found for energies $0 < \mc{E}_n \leq \egap$ and correspond to deformations of the skyrmion into polygons. From the explicit form of $\mc{H}_m$ given in Eq.~\ref{eq:MagnHam}, it becomes apparent that if $e^{i m \phi} \psi_{n,m}(\rho)$ is a magnetic excitation over the skyrmionic field ($\mu=1$), then $e^{-i m \phi} \psi_{n,m}(\rho)$ is an excitation over the antiskyrmionic field ($\mu=-1$), with the same energy. These two states have an opposite sense of gyration, $\sim \cos(\mc{E}_{n,m} t\pm m \phi)$, respectively. 

\begin{figure}[t!]
\centering
\includegraphics[width=1\columnwidth]{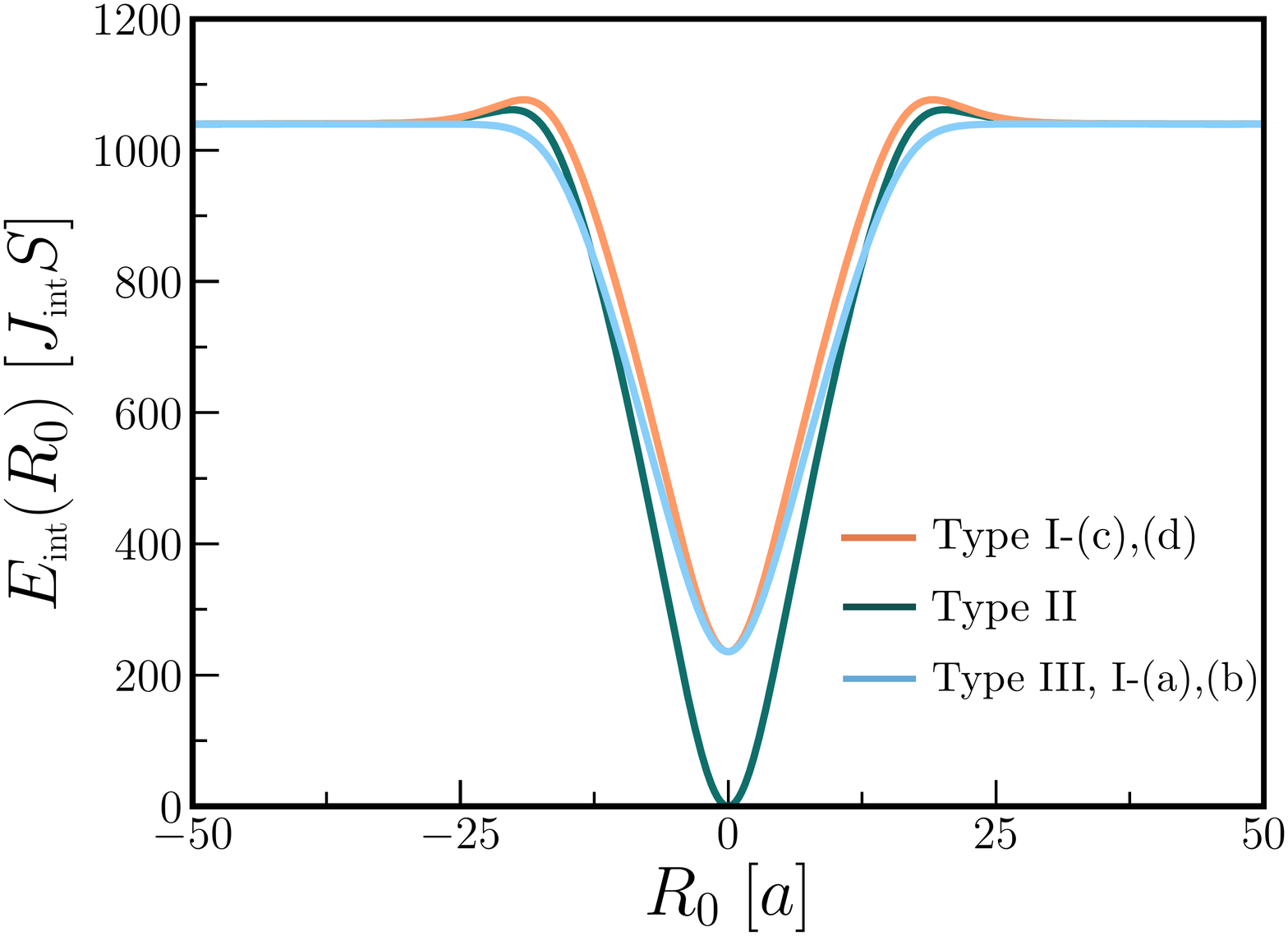}
\caption{Interaction potential as a function of the distance of the center-of-mass of the two particles. Topological textures with different helicities have the same interaction energy, irrespective of their topological charge (blue line). Textures with the same helicity interact via a potential that distinguishes between pairs with the same (green line) or opposite topological charge (orange line). Each topological particle is of size $8.8a$.}
\label{fig:inter}
\end{figure}
 
We now turn our attention to the calculation of the interaction potential $\Eint$, between the two topological particles. In the continuum limit, the bilayer interacting Hamiltonian equals $\Hint = -\Jint \int  \mb{n}_1(\mb{r}) \cdot \mb{n}_2(\mb{r}) d\mb{r}$. Since the skyrmion is a localized object, we can get an intuition of the interlayer interaction by introducing a set of collective coordinates as $\mb{n}_i (\mb{r})= \mb{n}_i(\mb{r}-\mb{R}_i)$, where here $\mb{R}_i$ represents the particle center of mass, and $i=1,2$ is the layer index. Thus, particles on different layers interact through a potential of the form $\Eint (R_0)=\Jint \int d\mb{r} [1-\mb{n}_1(\mb{r}-\mb{R}_1) \cdot \mb{n}_2(\mb{r}-\mb{R}_2)] d\mb{r} $, and $R_0=\vert \mb{R}_1 -\mb{R}_2\vert$ \cite{Koshibae2017}. For reasons of simplicity, in all considered cases, the $J$ and $D$ couplings in both layers have the same strength, thus the skyrmion and antiskyrmion have the same size, which is determined by the competition among the Heisenberg, DM, and Zeeman interaction. The behavior of $\Eint$, depicted in Fig.~\ref{fig:inter}, depends on both the helicity and the topological charge of the composite pair. Topological textures with different helicities have the same interaction energy, irrespective of their topological charge (blue line). Textures with the same helicity interact via a potential that distinguishes between pairs with the same (green line) or opposite topological charge (orange line).

As a final note, we derive the equation of motion for the collective coordinate of the topological charge separation for the Type I bilayer. Employing Thiele's approach \cite{PhysRevLett.30.230} in the limit $Q \gg \alpha$, we obtain the equation of motion for the collective coordinates of position of each layer $\mb{R}_i$,
\begin{align}\label{eq:Thiele}
-Q_i \epsilon_{ \nu \mu} \dot{R}_i^{\mu} = \frac{\pt \Eint}{\pt R_{i}^{\nu}} + f_i^{\nu}(t) \,,
\end{align}
where $i=1,2$ is the layer index, $\mu,\nu = x,y$, $\epsilon_{\mu \nu}$ is the antisymmetric tensor, $\Eint$ is the interaction energy due to the interlayer coupling, and $f_i^{\nu}(t)$ is a time-periodic function of frequency $\omega$, which parametrizes the interaction between the coordinate $R_i^{\nu}$ and the gyrotropic mode activated by the in-plane time-periodic magnetic field. For the special case of the Type I bilayer, we note that Eq.~\eqref{eq:Thiele} is considerably simplified by making use of the fact that the two layers host particles with opposite topological charge, $Q_1=-Q_2$, and the property $ \pt \Eint/\pt R_{1}^{\nu} = - \pt \Eint/\pt R_{2}^{\nu}$. Thus, the equation of motion for $\mb{R}_0=\mb{R}_1-\mb{R}_2$ takes the form $-Q \epsilon_{\nu \mu} \dot{R}_0^{\mu} = f_0^{\nu}(t)$, with $Q=Q_1$. In view of the numerical results for $\mb{R}_0$, we use the ansatz $f_0^{\nu}(t) = F_{\nu}(\tilde{\gamma}) \cos(\omega t)$, with $F_x(\tilde{\gamma})= c \sin(\tilde{\gamma}/2)$, $F_y(\tilde{\gamma})= c \cos(\tilde{\gamma}/2)$, and $c$ a constant, allowing for a dependence on the helicity difference $\tilde{\gamma} = \gamma_1 - \gamma_2$.

\end{document}